\documentclass[11pt]{article}

\usepackage{amsmath}
\usepackage{geometry} 
\usepackage{graphicx} 
\usepackage{color} 
\usepackage{amssymb, amsthm} 
\usepackage[]{natbib}
\usepackage{hyperref} 

\usepackage{booktabs}
\usepackage[table,xcdraw,dvipsnames]{xcolor}
\usepackage{soul}
\usepackage{caption}
\usepackage{subcaption}
\usepackage{lscape}
\usepackage{setspace}
\usepackage{setspace}
\usepackage{float}
\usepackage{comment}

\usepackage[title]{appendix}

\newtheorem*{remark}{Remark}

\newcommand{\blind}{0}

\title{Hypergraph adjusted plus-minus}

\if1\blind
{
  \author{[Blinded for review]}
}\fi

\if0\blind
{
\author{Nathaniel Josephs$^1$ and Elizabeth Upton$^2$ \\[1em]
\normalsize 1. North Carolina State University, Department of Statistics \\
\normalsize 2. Williams College, Department of Mathematics and Statistics}
}\fi

\date{}

\begin{document}
\maketitle

\begin{abstract}

In team sports, traditional ranking statistics do not allow for the simultaneous evaluation of both individuals and combinations of players.  Metrics for individual player rankings often fail to include the interaction effects between groups of players, while methods for assessing full lineups cannot be used to identify the value of lower-order combinations of players (pairs, trios, etc.).  Given that player and lineup rankings are inherently dependent on each other, these limitations may affect the accuracy of performance evaluations.  To address this, we propose a novel adjusted plus-minus (APM) approach that allows for the simultaneous ranking of individual players, lower-order combinations of players, and full lineups within a team.  The method adjusts for the complete dependency structure and is motivated by the connection between APM and the hypergraph representation of a team.  We discuss the similarities of our approach to other advanced metrics, demonstrate it using NBA data from 2012-2022, and suggest potential directions for future work.

\noindent 
\\[1em]
\end{abstract}

\section{Introduction}

Individual player rankings are central to the world of sports:
fans love to debate whether their favorite players are currently or historically the best, those same players are incentivized to perform due to contract negotiations and season-end awards like Most Valuable Player, and team management is invested in constructing a winning roster.
Player rankings serve as a foundation for healthy competition, strategic decision-making, and fan engagement, underscoring the importance of their accuracy for everyone involved.  For individual sports, rankings can be obtained easily through time trials and head-to-head matchup results. 
But for team sports, disentangling individual effects from endogenous factors like team synergy and opponent, as well as exogenous factors like coaching and home-field advantage, can be a challenging exercise.  

One of the most common approaches to measuring an \textit{individual's} effect in team sports is raw plus-minus (PM).
Originally used in hockey, a player's PM is their team's cumulative difference in score when that player is in the game.
While intuitively appealing, PM is sensitive to substitution patterns.
For instance, one very good player can artificially inflate the PM of their teammates, and quantifying this effect becomes challenging when certain combinations of players consistently share playing time. 
The most common approach to account for this is known as adjusted PM (APM).
APM was developed for basketball
\citep{rosenbaum2004measuring} and hockey \citep{macdonald2011regression} as a method for comparing individual players while accounting for those that play together.
It has also been extended, via a Bayesian framework, to soccer and American football \citep{MatanoRichardsonPospisilPolitschQin+2023+43+49, sabin2021estimating}.  
We provide some mathematical details of APM in Section~\ref{sec:APM}.

The intention of APM and its variants is to isolate an individual player's effect \citep{hvattum2019comprehensive}.
However, this approach fails to capture any \textit{interaction} effects between players.
Experience, complementary skill sets, and general synergy often result in a group of players whose combined effect surpasses the sum of their individual contributions.
This phenomena is in part why substitution patterns are not uniformly random, and why managers and coaches play a crucial role in the success of a team.  Likewise, when assessing and ranking an individual's contribution to a team's success, it is critical to account for the player's ability to create synergies with the various combinations of their teammates.  We refer to these various combinations (individuals, pairs, trios, etc.) as \textit{generalized lineups} and propose a novel approach to quantify the contribution of all generalized lineups for a team simultaneously. 

To evaluate generalized lineups using APM, there are two possible options: 
\begin{enumerate}
    \item Sum the APM of all individual players in the generalized lineup of interest. This ignores all interaction effects and assumes that the performance of the generalized lineup is equivalent to the sum of the individual performances of the players in the lineup. 
    \item Find the cumulative PM for the generalized lineup, i.e. find the score differential while that combination of players is in the game. This does not incorporate any lineup adjustment and is sensitive to substitution patterns in the same manner as the simple PM statistic.
\end{enumerate}

Both of these approaches have obvious drawbacks, and neither is able to rank the generalized lineups while accounting for interaction effects.
To fill the gap between these two extremes, we propose a solution for ranking lineups and players that adjusts not only for a full lineup, but all lower-order combinations of players.
Our methods are motivated by the hypergraph representation of a lineup, which we show is a generalization of the design matrix used in the ordinary APM method.
The result is a unified approach for simultaneously ranking any generalized lineup including individuals, pairs, and full lineups.  Furthermore, our methods lend themselves to interpretable visualizations that allow one to quickly identify positive and negative synergies among groups of players on a team.   

The rest of the paper is organized as follows.
In Section~\ref{sec:related}, we review related methods in the literature for ranking lineups.
We then provide the mathematical background for APM, used to rank individual players, in Section~\ref{sec:APM}.
In doing so, we discuss pairs APM (a variant of APM that can be used to rank and adjust for pairs of players, but is ultimately too computationally expensive for higher-order generalized lineups).
In Section~\ref{sec:hypergraphs}, we review the necessary graph theory terminology and notation, and then make the connection between APM and hypergraphs.  This leads to the development of our two new ranking methods.  We apply these new methods to 10 years of NBA data from 2012-2022 in Section~\ref{sec:analysis} and conclude with a discussion in Section~\ref{sec:discussion} of our methods, their limitations, and potential future directions.

\section{Related Work}\label{sec:related}

While methods for ranking lower-order combinations of players and full lineups simultaneously are lacking, there has been a great deal of research on measuring the efficiency of full lineups.  For example,
\cite{lechner2012superior} take a resource-based view for choosing optimal lineups in soccer and describe the type of ``human resources" that lead to superior value creation on a team.
\cite{maymin2013nba} introduce a skills plus-minus to measure how good individuals are at certain aspects of basketball such as ballhandling, rebounding, and scoring.
The authors then compute optimal lineups by measuring synergies and finding players with complementary skill sets.
\cite{grassetti2021extended} discuss how APM can be used directly for ranking full lineups: Besides summing over individual player effects in a given lineup, the authors propose a separate parameter for each lineup, constituting a lineup effect.
\cite{barrientos2023bayesian} introduce a Bayesian method for quantifying the high degree of uncertainty in individual and lineup rankings.  

There are a few existing ranking methods that use techniques from network theory.
LinNet was introduced in \cite{pelechrinis2018linnet} as an approach to ranking lineups via network embeddings.
A network is constructed in which the lineups are nodes, a directed edge $i \rightarrow j$ exists if lineup $i$ outscored lineup $j$, and the corresponding scoring margin is the weight of the edge.
For each lineup, the authors compute the average APM of the players in the lineup.
The authors then embed this lineup network using \texttt{node2vec} and the resulting latent vectors are used as covariates in logistic regression.
The regression response is the difference in average APM for two nodes, which allows the authors to infer the probability that one lineup outperforms another.  Similarly, \cite{ahmadalinezhad2020basketball} use lineup networks to predict the matchup result between opposing lineups.
The authors introduce Inverse Square Metric as a measure for lineup performance prediction, with $\text{ISM}(i, j)~=~\text{D}(i)\text{D}(j) / \vert \text{LP}(i, j) \vert^2$, where $\text{D}(i)$ is the degree of lineup $i$ and $\text{LP}(i, j)$ is the length of the shortest path between lineups $i$ and $j$.
The authors demonstrate that ISM is more predictive of performance than the \texttt{node2vec} embeddings.

While the aforementioned approaches implicitly recognize that evaluating a lineup is not as simple as summing the contributions of the individual players within the lineup, they do not account for the complete scope of the interaction effects.  Furthermore, they cannot be used for ranking generalized lineups outside of full lineups.

The method most similar to ours is \cite{devlin2020identifying}, which takes a spectral approach.
The authors define a \textit{success function} $f(L) = \text{pm}_L$, where $L$ is one of the $\binom{15}{5}$ 5-person lineups from a roster of 15 and $\text{pm}_L$ is the lineup's raw plus-minus.
The success function is shown to decompose orthogonally into the generalized lineups, hence providing a natural measure of contribution from each generalized lineup. 
Inference is performed by relating the success function to the \textit{Johnson graph}.
To account for the number of possessions a lineup plays together, the authors normalize their results by the log of the number of possessions played by each lineup, which they call \textit{spectral contribution per log possession (SCLP)}. There is not a natural means to adjusting for opponent with SCLP.  

In contrast to SCLP, we utilize a model-based approach via the familiar tools of regression analysis.
Our results, as fitted values or coefficient estimates, are easy to interpret and can provide standard error estimates and confidence statements.  Furthermore, whereas SCLP requires data from a stable 15-person roster, our methods can be utilized to make predictions for unused lineups and can consider teammates that are added, dropped, or traded throughout the season.  
While both metrics provide within team rankings, we show that
our results nevertheless provide an accurate comparison of generalized lineups across the league.
Lastly, as noted by the authors of~\cite{devlin2020identifying}, the various methods of rating players and generalized lineups can be seen as complementary: Each approach offers distinct insights, and when combined, they enhance decision-making by providing a more comprehensive understanding of team and player performance.


\section{Adjusted Plus-Minus}\label{sec:APM}

To isolate and rank individual players' effects, adjusted plus-minus (APM) utilizes multiple linear regression.
Let $X$ be an $n \times p$ matrix where $n$ is the number of unique lineups over the time period considered, $p$ is the number of unique players, and $X_{ij} = 1$ if player $j$ was in lineup $i$ and 0 otherwise.
The response variable $Y$ is the on-court differential (points scored - points allowed) for that lineup.
Assuming a linear model where $\mathbb{E}[Y|X] = X\beta$, then via ordinary least squares, APM is the the vector of coefficient values
$\widehat{\beta_{ols}} = (X^TX)^{-1}X^TY$ from the fitted model.
Hence, $\widehat{\beta_{ols}}$ represents the contribution of each of the $p$ players, given the other players on the team, and can be used to rank player impact.

To illustrate APM, we introduce a hypothetical example that will be referred to throughout the manuscript.  Imagine a team with 5 players labeled A-E, and a 10-minute game with lineups of size 3.
The results of one such game are summarized in Table~\ref{tab:lineup}. 
In the linear model described above, the player columns (A:E) define the design matrix $X$ and `Score Dif' is the response vector $Y$. 
We see that the first lineup (A, B, C) was the most successful with a score differential of +3, indicating that this lineup outscored the other team by 3 total points in the 3 minutes that A, B, and C played together.
We also see that Player A has the top ranking PM score of +5, which can be computed via the dot product $X_{\cdot, A} \cdot Y$.
However, Player A always played with Player B, so it is unclear which of the two players deserves more credit for their lineups' successes.

\begin{table}[]
    \centering
    \begin{tabular}{ccccccc}
      Score Dif & Time (min) & A & B & C & D & E \\
      \hline
      3 & 3 & 1 & 1 & 1 & 0 & 0 \\
      2 & 3 & 1 & 1 & 0 & 1 & 0 \\
      -1 & 2 & 0 & 0 & 1 & 1 & 1 \\
      -2 & 1 & 0 & 1 & 0 & 1 & 1 \\
      0 & 1 & 1 & 1 & 0 & 0 & 1 \\
    \end{tabular}
    \caption{Toy example for a 10-minute game of 3-on-3 basketball. The table concatenates the response vector, diagonal of time matrix, and design matrix, i.e. $[Y,~\text{diag}(W),~X]$.}
    \label{tab:lineup}
\end{table}

In our example, the fitted coefficients from regressing $Y$ on $X$, shown in Table~\ref{tab:coefficients}, result in the following player rankings (in order from best to worst): A, C, B = D, E.
This model fits the data perfectly given $n = p = 5$, and the coefficient estimates are very sensitive to small perturbations in the data.
Furthermore, the existence of rank ties, which we see with players D and E, is undesirable.  More advanced APM metrics incorporate both home and away players (i.e. each unique observation is a lineup of one
team vs the other team), use a penalized regression setup to stabilize the estimates~\citep{sill2010improved}, and incorporate playing time and additional box-score statistics or external ratings~\citep{hvattum2019comprehensive}.  In our example, after using cross-validation to choose an appropriate level of regularization, a weighted ridge regression results in the improved ranking scheme: A, C, B, D, E (see $\widehat{\beta_r}$ in Table~\ref{tab:coefficients}).  Alternatively, if information was known \textit{a priori} about our example players, we could recast the ridge regression into a Bayesian framework and incorporate this additional player information in the prior distribution~\citep{MatanoRichardsonPospisilPolitschQin+2023+43+49}.  



\begin{table}[]
    \centering
    \begin{tabular}{c|cccc}
      Player & $\text{PM}$ &$\widehat{\beta}_{ols}$ & $\widehat{\beta}_r$ & $\widehat{\beta}_\text{PAPM}$ \\
      \hline
      A & 5 & 2 & 1.64 & 0.254 \\
      B & 3 & 0 & 0.35 & 0.253 \\
      C & 2 & 1 & 0.90 & 0.101 \\
      D & -1 & 0 & -0.12 & -0.005 \\
      E & -3 & -2 & -1.78 & -0.108 \\
    \end{tabular}
    \caption{PM, APM using OLS, APM using ridge regression, and PAPM for players A-E using the data from Table~\ref{tab:lineup}.}
    \label{tab:coefficients}
\end{table}

While a popular player-ranking metric, a shortcoming of APM and its variants is that it does not adjust for generalized lineups.
Consider the scenario where two players excel when paired together but perform less effectively when playing separately. APM fails to capture this nuance, and may result in both individual players being ranked close to the median depending on their frequency of playing together versus apart.
On the other hand, some individually great players may not play well together if, for example, their skill sets overlap.
It is therefore necessary to account for combinations of players within full lineups.

\subsection{Pairs APM}

An intuitive and relatively straightforward approach to account for lower-order combinations in the current construction of APM is to add interaction terms as new columns in the design matrix.  Continuing with our toy example, we can add $\binom{5}{2} = 10$ additional columns to X, each representing a pair of players.
The column vector takes the value 1 when both players are in the line-up and 0 otherwise.
See Table~\ref{tab:papm} for an example.
The response vector does not change, and the coefficients obtained from regressing $Y$ on $X$ allow one to simultaneously rank individuals and pairs while accounting for all other players and pairs. 
We refer to this method as pairs adjusted plus-minus (PAPM).

We include the coefficients from PAPM in Table~\ref{tab:coefficients} found using weighted ridge regression.
With a new estimation procedure, the individual player rankings can change compared to the $\hat{\beta_r}$ estimates, e.g. player B now outranks player C (although this case may be due to the fact that, as a consequence of ridge regression, player B receives a larger contribution simply by virtue of appearing in four of the five lineups).
Furthermore, there is the added benefit of ranking pairs by utilizing the regression coefficients associated with the additional columns in the design matrix.  For context, the pair rankings from best to worst are AB, BC, AC, AD, BD, and so on.

\begin{table}[H]
    \centering
    \begin{tabular}{ccccccccccccccccc}
      Score Dif & Time (min) & A & B & C & D & E & AB & AC & AD & AE & BC & BD & BE & CD & CE & DE \\
      \hline
      3 & 3 & 1 & 1 & 1 & 0 & 0 & 1 & 1 & 0 & 0 & 1 & 0 & 0 & 0 & 0 & 0 \\
      2 & 3 & 1 & 1 & 0 & 1 & 0 & 1 & 0 & 1 & 0 & 0 & 1 & 0 & 0 & 0 & 0 \\
      -1 & 2 & 0 & 0 & 1 & 1 & 1 & 0 & 0 & 0 & 0 & 0 & 0  & 0 & 1 & 1 & 1 \\
      -2 & 1 & 0 & 1 & 0 & 1 & 1 & 0 & 0 & 0 & 0 & 0 & 1 & 1 & 0 & 0 & 1 \\
      0 & 1 & 1 & 1 & 0 & 0 & 1 & 1 & 0 & 0 & 1 & 0 & 0 & 1 & 0 & 0 & 0 \\
    \end{tabular}
    \caption{Table~\ref{tab:lineup} with additional columns for player pairs. Note that only pairs with non-zero playing time are computed, i.e. the column sums for each pair are all positive.}
    \label{tab:papm}
\end{table}

We note that this idea was alluded to in~\cite{macdonald2011regression}:
\begin{quote}
    \textit{By not including interaction terms in the model, we do not account for interactions between players. Chemistry between two particular teammates, for example, is ignored in the model. The inclusion of interaction terms could reduce the errors. The disadvantages of this type of regression would be that it is much more computationally intensive, and the results would be harder to interpret.}
  \end{quote}
As mentioned, the immediate drawback to this approach is its scalability.  The number of possible combinations of players explodes as we consider larger generalized lineups, resulting in a design matrix where ``$p >> n.$"  Regularization techniques can be used to make the regression computationally possible but the results are not tenable for interaction terms  composed of more than two players.
Even so, we explore this method in Section~\ref{sec:analysis} when applying our proposed methods to historical NBA data.

\section{Hypergraphs}\label{sec:hypergraphs}

In this section, we introduce networks, hypergraphs, and line graphs, which are essential to our proposed methods.
We begin by connecting the design matrix of APM to a $k$-regular hypergraph.
We then introduce a generalization of this to capture all generalized lineups.
Finally, we describe our two new methods.

\subsection{Background}\label{sec:background}

Graphs, or networks, are relational data objects composed of nodes (also known as vertices) that are linked by edges.
Edges represent various types of relationships or interactions between nodes, and a variety of statistical methods can be utilized to model and understand the complex dependencies captured by the network \citep{kolaczyk2014statistical}.
Given that team lineups are inherently interconnected, networks of game/team data are a natural representation to consider when ranking generalized lineups.

\begin{figure}[h]
    \centering
    \begin{minipage}{0.39\textwidth}
        \centering
        \[
       \begin{array}{r@{}c}
    \begin{tabular}{c|lllll}
          & \multicolumn{5}{l}{\textbf{Edges(Lineups)}} \\
\shortstack{\textbf{Nodes}\\\textbf{(Players)}} & R & Y & G & B & P \\
\hline
A & 1 & 1 & 0 & 0 & 1 \\
B & 1 & 1 & 0 & 1 & 1 \\
C & 1 & 0 & 1 & 0 & 0 \\
D & 0 & 1 & 1 & 1 & 0 \\
E & 0 & 0 & 1 & 1 & 1   
\end{tabular}
\end{array}
        \]
        \label{tab:incidence}
    \end{minipage}%
    \hfill
    \begin{minipage}{0.3\textwidth}
        \centering
        \includegraphics[width=\textwidth]{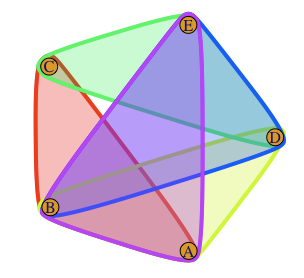}
    \end{minipage}
        \hfill
    \begin{minipage}{0.3\textwidth}
        \centering
        \includegraphics[width=\textwidth]{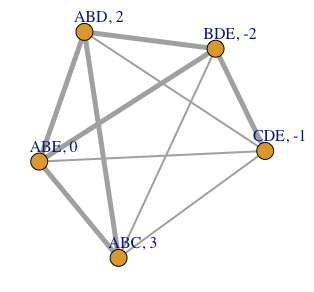}
    \end{minipage}
     \caption{The incidence matrix $M$ (left) for the hypergraph (middle) is exactly the transpose of the design matrix $X$ for APM given in Table~\ref{tab:lineup}. The hyperedges are colored to enhance the visualization: R (red), Y (yellow), G (green), B (blue), and P (purple). The line graph (right) of the hypergraph (middle) is formed by swapping the roles of nodes and hyperedges.}
    \label{fig:incidence}
\end{figure}

Mathematically, a network is a graph $G = (V, E)$ with node set $V = \{1,2,3,..., n\}$ and edge set $E~\subset~V \times V.$ 
Networks are often represented by an adjacency matrix $A \in \{0,1\}^{n \times n}$ where $A_{ij}~=~1$ whenever $(i,j) \in E$ and $0$ otherwise.
Other important matrix representations include:
\begin{itemize}
    \item the graph Laplacian: $L = D - A$, where $D$ is the diagonal degree matrix, i.e. $D_{ii} = \sum_{j \in V} A_{ij}$
    \item the incidence matrix: $M,$ which has a row for each node, a column for each edge in the network, and $M_{ij} = 1$ if node $i$ is incident to edge $e_j.$
\end{itemize}  
A hypergraph extends the concept of a graph by allowing edges, called hyperedges, between more than two nodes.  Hypergraphs are a natural representation for team lineups, where the nodes are players, and hyperedges represent lineups.  The hypergraph visualization for our toy example is in the center of Figure~\ref{fig:incidence}.  The nodes represent the players A-E, and are connected to each other by hyperedges that represent the lineups. We have colored the hyperedges for a clearer visualization, e.g. red (R) represents the lineup A-B-C, which appears in the first column of Table~\ref{tab:lineup}.

Importantly, we realize the design matrix in Table~\ref{tab:lineup} as the transpose of the incidence matrix of its hypergraph.  Figure~\ref{fig:incidence} illustrates this connection between the design matrix and the hypergraph.  This connection between APM and hypergraphs motivates a deeper exploration into the possibility of a network analysis to inform the ranking of players and generalized lineups.

Statistical methods that incorporate network information through hypergraphs are less developed than those for standard graphs. A common approach to integrating the hypergraph structure into analysis is to transform the hypergraph into a standard graph. One such transformation is the line graph of the hypergraph.

A line graph is constructed by treating each hyperedge in the original hypergraph as a node in the new graph. Two nodes in the line graph are connected by an edge if their corresponding hyperedges in the hypergraph share at least one node. This transformation allows us to analyze relationships between hyperedges in a more conventional graph-based framework.  To quantify the strength of these connections, we assign weights to the edges in the line graph. Specifically, we use Jaccard similarity to measure the overlap between two hyperedges. The Jaccard similarity between two hyperedges $e_i$ and $e_j$ is given by:
\begin{equation*}
    w_{ij} = \frac{|e_i \cap e_j|}{|e_i \cup e_j|} \enskip ,
\end{equation*}
where $e_i$ and $e_j$ are the set of nodes the belong to the corresponding hyperedges.  For example, in our toy dataset (illustrated in Figure~\ref{fig:incidence}), consider the hyperedges ABD (yellow) and ABE (purple) in the hypergraph. In the corresponding line graph, these hyperedges become nodes, which are connected by an edge. The weight of this edge is calculated using Jaccard similarity: out of the combined set of 4 players A, B, D and E, players A and B are common to both nodes, resulting in an edge weight of $\frac{2}{4}.$  This weighting scheme reflects the degree of similarity between different hyperedges. Since we are working with a weighted graph, we replace the binary adjacency matrix with its weighted counterpart,  such that $A_{ij} = w_{ij}$ whenever $(i,j) \in E$.  Furthermore, note that by utilizing the line graph representation, the score differentials, $Y,$ are now associated with the \textit{nodes} in the line graph.

\begin{figure}[H]
  \begin{minipage}{0.5\textwidth}
    \centering
    \includegraphics[width=.8\linewidth]{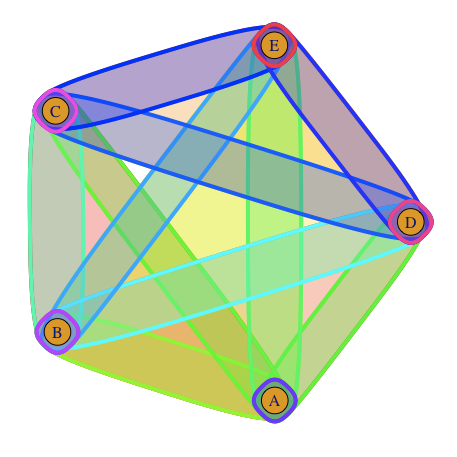}  
  \end{minipage}
  \begin{minipage}{0.5\textwidth}
    \centering
    \includegraphics[width=.8\linewidth]{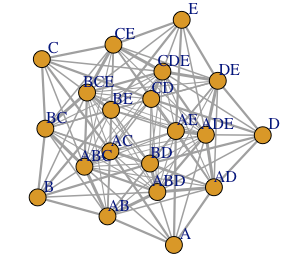}  
  \end{minipage}
  \caption{The extended hypergraph (left) defined by Table~\ref{tab:extended} and its corresponding line graph (right).}
  \label{fig:hypergraph_extended}
\end{figure}

The hypergraph and its corresponding line graph, as currently described, are based solely on full lineups.
This is known as a $k$-regular hypergraph, where each hyperedge connects exactly $k = 3$ nodes.
However, we can expand our representation to include lower-order player combinations by introducing additional hyperedges.  We refer to this as the \textit{extended hypergraph}, and the transpose of its incidence matrix as the \textit{extended APM design matrix}. Continuing with our toy example, we add additional rows to Table~\ref{tab:lineup} describing the aggregate pair and solo player results, as shown in the case of Player A, in Table~\ref{tab:extended}.  Note that we do not include redundant rows in the matrix, i.e. the A-B pairing and score differential is not duplicated in the extended lineup data for the B-A pairing.  We also create the line graph corresponding to the extended hypergraph; both are shown in 
Figure~\ref{fig:hypergraph_extended}.
In Figure~\ref{fig:hypergraph_extended}, the line graph (right) has a node for each generalized lineup, and the generalized lineups are connected via a weighted edge based on Jaccard similarity.

\begin{table}[H]
    \centering
    \begin{tabular}{ccccccc}
      Score Dif & Time (min) & A & B & C & D & E \\
      \hline
       \vdots & \vdots & \vdots & \vdots & \vdots & \vdots & \vdots \\
      5 & 7 & 1 & 1 & 0 & 0 & 0 \\
     3 & 3 & 1 & 0 & 1 & 0 & 0 \\
      2 & 3 & 1 & 0 & 0 & 1 & 0 \\
      0 & 1 & 1 & 0 & 0 & 0 & 1 \\
      5 & 7 & 1 & 0 & 0 & 0 & 0 \\
       \vdots & \vdots & \vdots & \vdots & \vdots & \vdots & \vdots \\
    \end{tabular}
    \caption{Extended lineup data from Table~\ref{tab:lineup} with generalized lineups. Here, we only show the additional rows corresponding to all lineups containing player A.}
    \label{tab:extended}
\end{table}

\subsection{Methods}\label{sec:methods}

Here, we present our two new methods that rank generalized lineups that are adjusted for the full lineup and all lower-order combinations of players: the Hypergraph Adjusted Plus-Minus (HAPM) and the Line Graph Adjusted Plus-Minus (LAPM).  In short, HAPM is regression using the extended APM design matrix and LAPM employs Bayesian regression using the line graph of the extended hypergraph with a prior that enforces smoothness on the fitted values of similar lineups.

\subsubsection{Hypergraph APM}

As an alternative to PAPM that is scalable for all generalized lineups, we introduce HAPM.
HAPM is the same weighted ridge regression described in Section~\ref{sec:APM} for APM, but with the original design matrix replaced by the transpose of the incidence matrix of the \textit{extended} hypergraph.
In this case, the columns of the design matrix represent each player, the rows are combinations of players, and the entry is 1 if that player is in that row's combination of players.
That is, $X_\text{HAPM}$ is the new design matrix with $({X_\text{HAPM}})_{ij} = 1$ if player $j$ is in the combination of players represented in row $i$ and 0 otherwise.
Note that if we restrict the rows to full lineups only, then this is exactly ordinary APM (calculated at the team level) and the design matrix is the transpose of the incidence matrix of the $k$-regular (non-extended) hypergraph.
As before, the response is the raw plus-minus for a given row.
Using ridge regression, we obtain coefficients $\widehat{\beta}_\text{HAPM}$ at the individual player level that account for all generalized lineups included in the design matrix.

Using HAPM, can we rank combinations of players?
The answer is yes.
In addition to the regression coefficients, which correspond to the players/columns of our design matrix, we have the predicted values $\widehat{y}_\text{HAPM}$ that correspond to the combinations/rows.
Therefore, we have a natural solution to rank combinations of any size.
For example, if we want to find the best combination of $m$ players, we sort the predicted values $\widehat{y}_\text{HAPM}$ among all combinations whose row sums of $X_\text{HAPM}$ are equal to $m$.
We can use this same approach to separately rank pairs, trios, etc.

\begin{remark}
    For individuals, we can use either the coefficients $\widehat{\beta}$ or the predicted values $\widehat{y}$ corresponding to row sums of $X$ equal to 1.
    These values are equivalent because $\widehat{y} = X\widehat{\beta}$ and for an individual $i$, there is only one nonzero column, $X_{\cdot, i}$, corresponding to that individual, hence $\widehat{y}_i = X_{\cdot, i}\widehat{\beta} = \widehat \beta_i$.
\end{remark}

\begin{remark}
    We can similarly use the predicted values $\widehat{y}$ of an APM regression model to rank full lineups.
    However, APM cannot be used to obtain rankings for any lower-order combination of players.    
\end{remark}

\subsubsection{Line Graph APM}

HAPM utilizes the hypergraph representation implicitly, yet it does not explicitly consider the connectivity patterns within the hypergraph. To address this, we draw inspiration from functional data analysis~\citep{ramsay} and propose a Bayesian regression approach using the line graph of the extended hypergraph. This approach resembles kernel regression on a graph~\citep{smola}, allowing us to extend regression-based techniques to graph-indexed data, where conventional models and assumptions may not be applicable.  In our case, it allows for a ranking of individual players and generalized lineups while considering \textit{all} active combinations of players on a team.

Utilizing the extended line graph representation of a team, we view the raw plus-minus associated with each node as a random response variable $Y_v$.
Our goal is to predict this value using information encoded in the network, and then rank the players and the generalized lineups based on these predictions.
We use a Bayesian regression model that is specified as follows:
\begin{align*}
    Y_v~|~\beta &\stackrel{\text{ind}}{\sim} N(\beta_v, ~\sigma^2 ) \\
    \beta &\sim N(0, ~\lambda^{-1}L_w^{-1})
\end{align*} 
where $\beta_v$ is a vertex indexed ``intercept," $\sigma$ is the standard deviation of the assumed normal distribution, $L_w$ is the weighted graph Laplacian introduced in section~\ref{sec:background}, and $\lambda$ is a regularization parameter.  Here, the maximum a posteriori (MAP) estimate for $\beta$ is
\begin{equation*}
    \hat{\beta}_\text{LAPM} = \textrm{argmax}_\beta \left\{\ell(\beta; Y) - \frac{\lambda}{2}\beta^TL_w\beta\right\}
\end{equation*}
showing that we seek estimates of $\beta_v$ that balance representativeness with respect to our observed data in the log-likelihood $\ell(\beta;Y)$ with prior information about the connectivity of the network captured by the Laplacian.
We refer to this as line graph adjusted plus-minus (LAPM).

Our choice of the graph Laplacian is designed to exploit information in the network to inform the model and force the predicted values to be smooth with respect to the topology of the line graph.  That is, nodes are assigned values based on their connectivity in the network and adjacent nodes will have similar predicted values.  In fact, the log prior can be expanded as
$$ \beta^TL_W\beta = \sum_{(i,j) \in E} w_{ij}(\beta_i - \beta_j)^2 $$
showing that the term penalizes the weighted sum of squared differences for coefficients of adjacent vertices in the network~\citep{kolaczyk2014statistical}.
Moreover, pairs of nodes with larger edge weights, corresponding to more similar generalized lineups, will be forced to have more similar predicted values.

Given that $L_w$ is a symmetric matrix, it realizes the eigendecomposition $L_w = \Phi\Xi\Phi$, and we can therefore represent $\beta$ using a basis expansion with respect to the eigenvectors of $L_w$ by letting $\beta = \Phi\theta$~\citep{upton}.
In short, we use the eigenvectors of the Laplacian as predictor variables, find a set of corresponding coefficients $\hat{\theta}$ through Bayesian linear regression, and then perform matrix-vector multiplication to solve for the vertex-indexed intercepts $\widehat{\beta_v} = \widehat{Y_v}.$  We do not have to use the full set of eigenvectors to represent $\beta$.
Instead, we can extract most of the information in $L_w$ through a low-order representation such that $\beta = \Phi_{1:\tau}\theta$, where $\Phi_{1:\tau}$ represents the first $\tau$ eigenvectors (ordered by the eigenvalues $\xi_1 < \dots <\xi_{\tau}$).  In practice, we select $\tau$ based on an `elbow' method, examining sequential differences in the size of ${\xi_i}$.


To gain some intuition regarding LAPM, consider player C in the line graph in Figure~\ref{fig:hypergraph_extended}.
Player C is represent by its own node, which is also connected to nodes BC, AC, CD, and CE (representing pairs) and ABC and CDE (representing full lineups) because player C is part of those generalized lineups.
Thus, the LAPM predicted value for C will be impacted, through the prior incorporating the Laplacian, by the predicted values/performance of all the aforementioned nodes.

Similarly to HAPM, we can rank player combinations of any given size using LAPM.  In this case, each vertex/node, corresponding to a generalized lineup, has its own coefficient $\widehat{\beta_v}$.
Coefficients corresponding to generalized lineups of the same size can be ordered and ranked. We can also use the line graph representation of the extended hypergraph to visualize the connectivity between players and the generalized lineups.

\section{NBA Analysis 2012-2022}\label{sec:analysis}

\subsection{Data Processing}

In this section, we apply PAPM, HAPM, and LAPM to NBA data from 2012-2022.
The raw play-by-play data can be purchased from \url{https://www.bigdataball.com/}.
Our code can be found at the GitHub repository \texttt{njosephs/HAPM}.

Our data processing pipeline is as follows.
For each team, we begin by computing the raw plus-minus (PM) data for each game.
We then aggregate PM over each season, i.e. we compute $X$, $W$, and $Y$ where the rows of $X$ are the unique lineups, $W$ is the average time per game played together for a given lineup, and $Y$ is that lineup's PM over the course of the season.
We restricted our analysis to only include players with at least 10,000 seconds of playing time, which is about 2 minutes per game.
For comparison, the minimum time for all-pro award eligibility was changed in 2023 to 65 games and 20 min/game, which is 78,000 seconds.
For simplicity, we do not adjust for the COVID shortened seasons (2019-2020 and 2020-2021), which, given the smaller sample size, will result in smaller between player variance for the APM statistics at the league level.

In order to compare our metrics to more commonly used versions of APM, we also compute $X$, $W$, and $Y$ across the entire league for each season rather than at the team level.  In this case, $X$ has a column for every player in the league and each ``lineup" is composed of ten players, across two teams, that share court time. We incorporate defense by letting the entries of $X$ be $\pm 1$ depending on whether the player is on the home or away team.
That is,
\begin{equation*}
    X_{ij} = \begin{cases}
        +1 & \text{if player}~i~\text{is in lineup}~j~\text{for the home team} \\
        -1 & \text{if player}~i~\text{is in lineup}~j~\text{for the away team} \\
        0 & \text{else}
    \end{cases}
\end{equation*}
in which each row of $X$ has exactly five entries of +1 and five entries of -1.
By setting $Y$ as the PM with respect to the home team, the negative entries of $X$ ensure that $Y_j < 0$ corresponds to an away lineup with a positive PM.
We refer to these metrics at the league level with defense as APM\_league and PAPM\_league, and at the team level without defense as APM and PAPM, respectively.

\begin{remark}
Our methods, HAPM and LAPM, are calculated at the team level and do not incorporate the opposing players, as this would require some notion of ``negative edges."  Consequently, the responses for HAPM and LAPM (as well as APM and PAPM) represent the season-level net points for a generalized lineup, without considering the opposing players on the court. In contrast, the responses for APM\_league and PAPM\_league account for both the season-level net points of a generalized lineup and the specific set of opponents. Despite this distinction, we can still make league-level comparisons by aggregating fitted values across teams, as described in Section~\ref{sec:rankcorr}.  
\end{remark}

For all of the ridge regression models (APM, PAPM, and HAPM), we chose the regularization parameter $\lambda$ that minimizes the cross-validation error, which is one of the default choice for the \texttt{glmnet} package \citep{hastie2021introduction}.

Finally, for the case study analysis, we explore the variability in the fitted values of HAPM and LAPM, along with the implied uncertainty in their rankings.
For HAPM, we bootstrap by sampling with replacement the rows of the design matrix.
For each bootstrap sample, we fit a ridge regression model and use the model to make predictions on all generalized lineups.
We examine and present the variation in the fitted HAPM values across the bootstrapped samples.
We also rank the fitted values within combinations of the same size (individuals, pairs, etc.), providing an estimate of rank uncertainty.
For LAPM, we sample from the posterior distributions of the coefficients utilizing a Metropolis-adjusted Langevin algorithm.
The details are provided in Appendix~\ref{sec:appendix2}.

\subsection{Rank Correlation}\label{sec:rankcorr}

Evaluating player rankings is ultimately qualitative since there is no notion of ground truth.
However, as a baseline comparison, we can see how closely our rankings are correlated with those based on advanced metrics such as box score plus-minus (BPM), value over replacement player (VORP), player efficiency rating (PER), and win shares per 48 minutes (WS.48) \citep{basketball-reference}.
Ideally, the individual rankings we obtain from our methods will be in general agreement with known methods designed to rank individuals, while also providing rankings for all combinations of players.

\begin{figure}[t!]
    \centering
    \includegraphics[width=.4\textwidth]{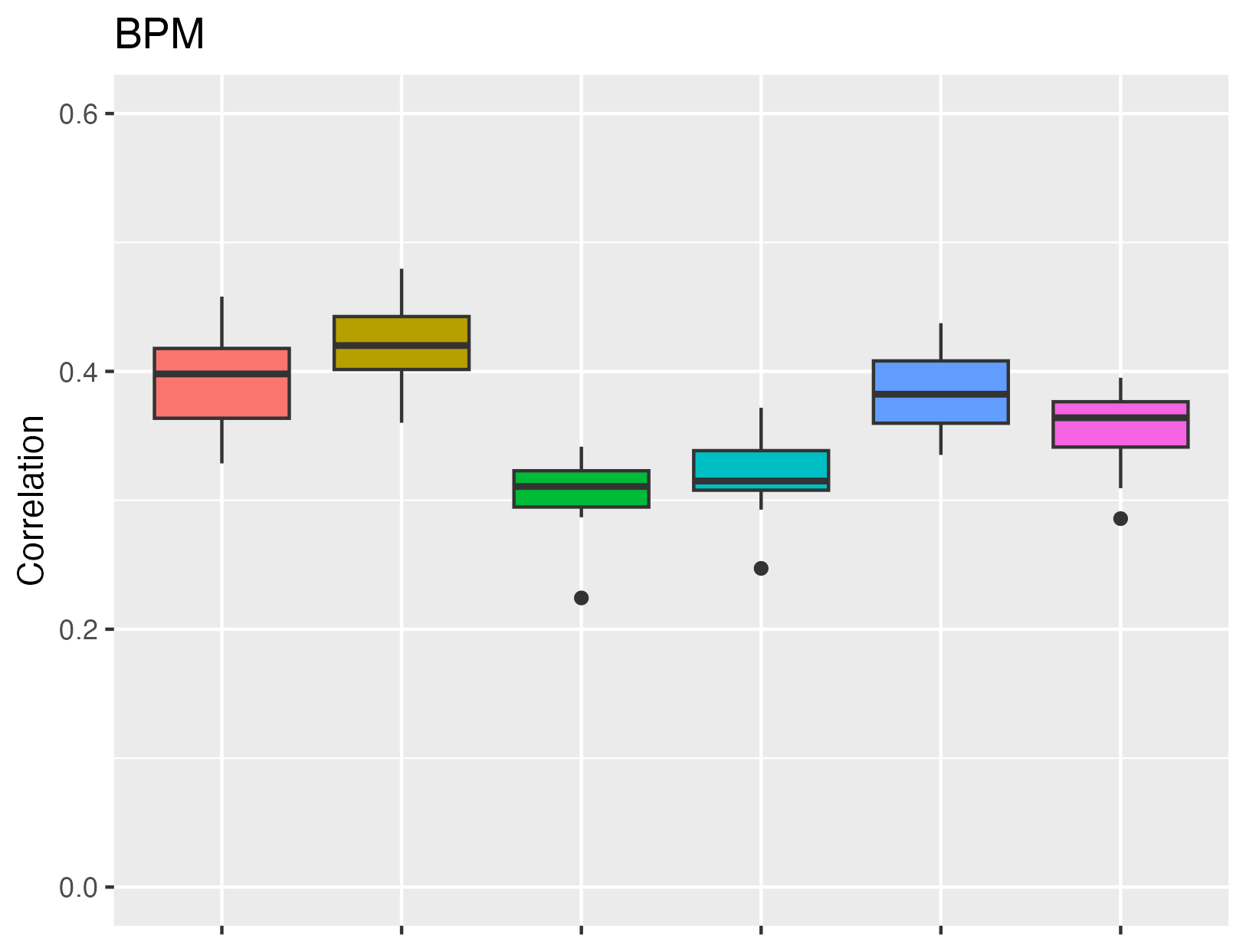}
    \includegraphics[width=.4\textwidth]{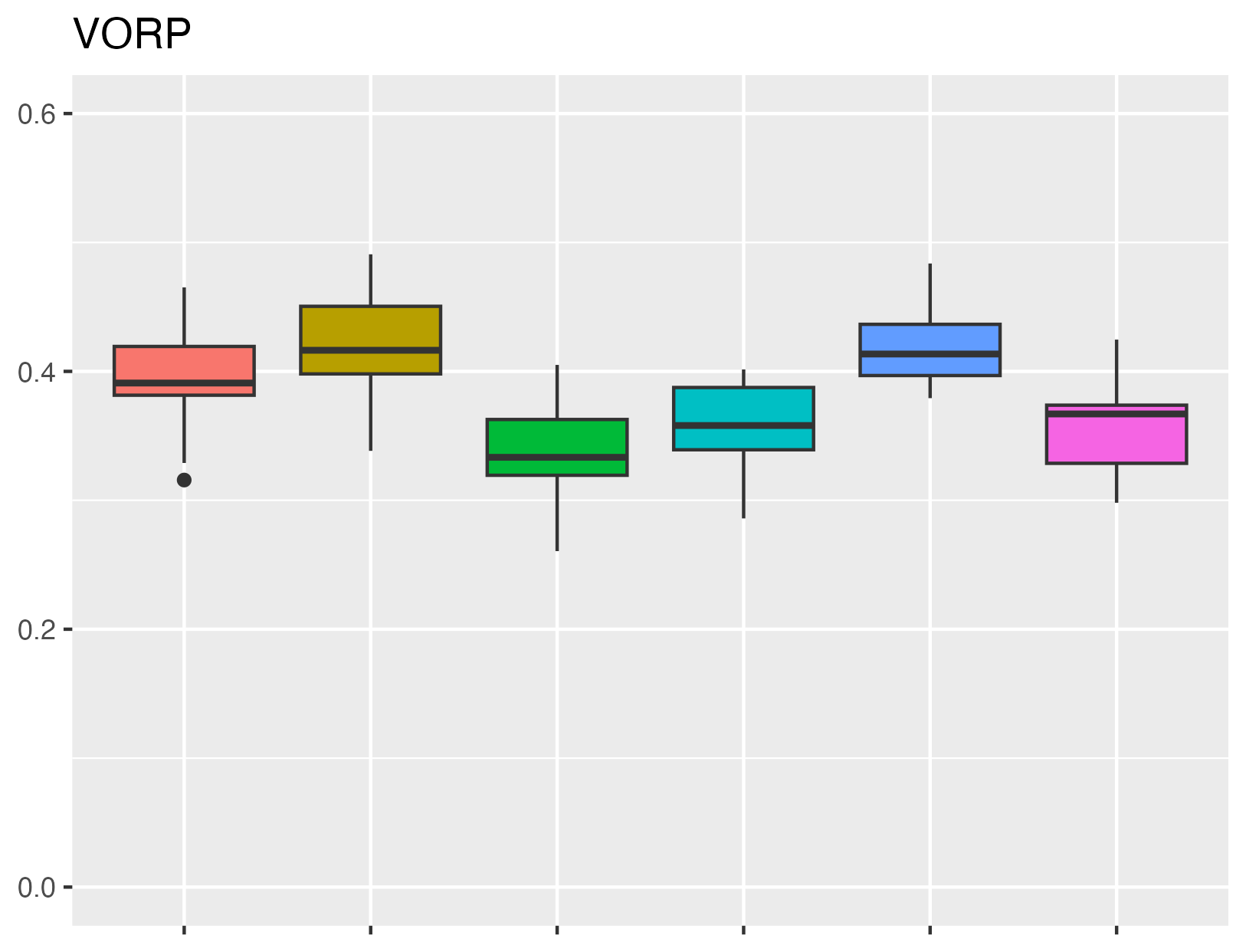}
    \includegraphics[scale = .25]{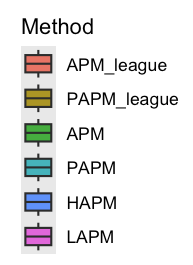}
    \hspace*{-2cm}  
    \includegraphics[width=.4\textwidth]{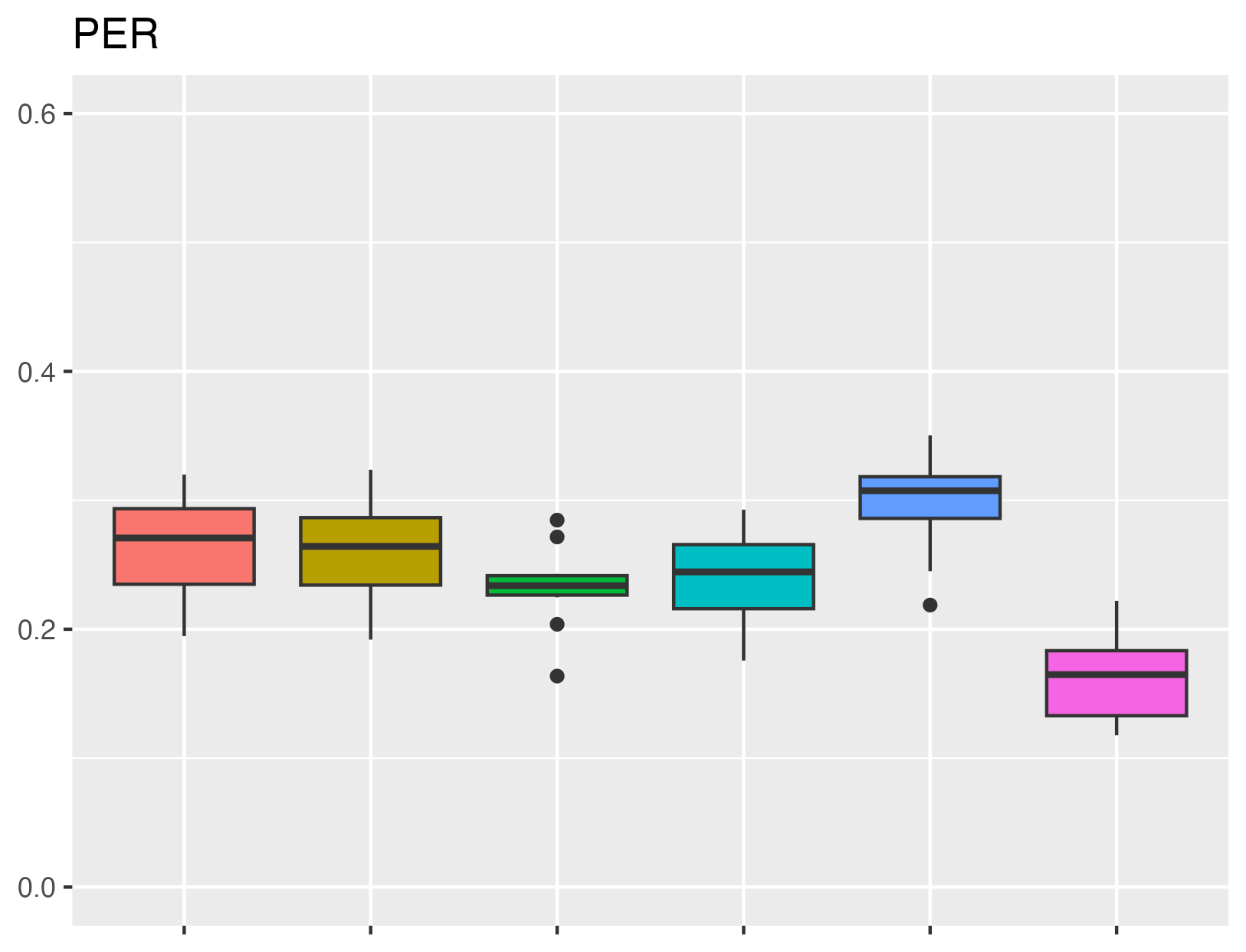}
    \includegraphics[width=.4\textwidth]{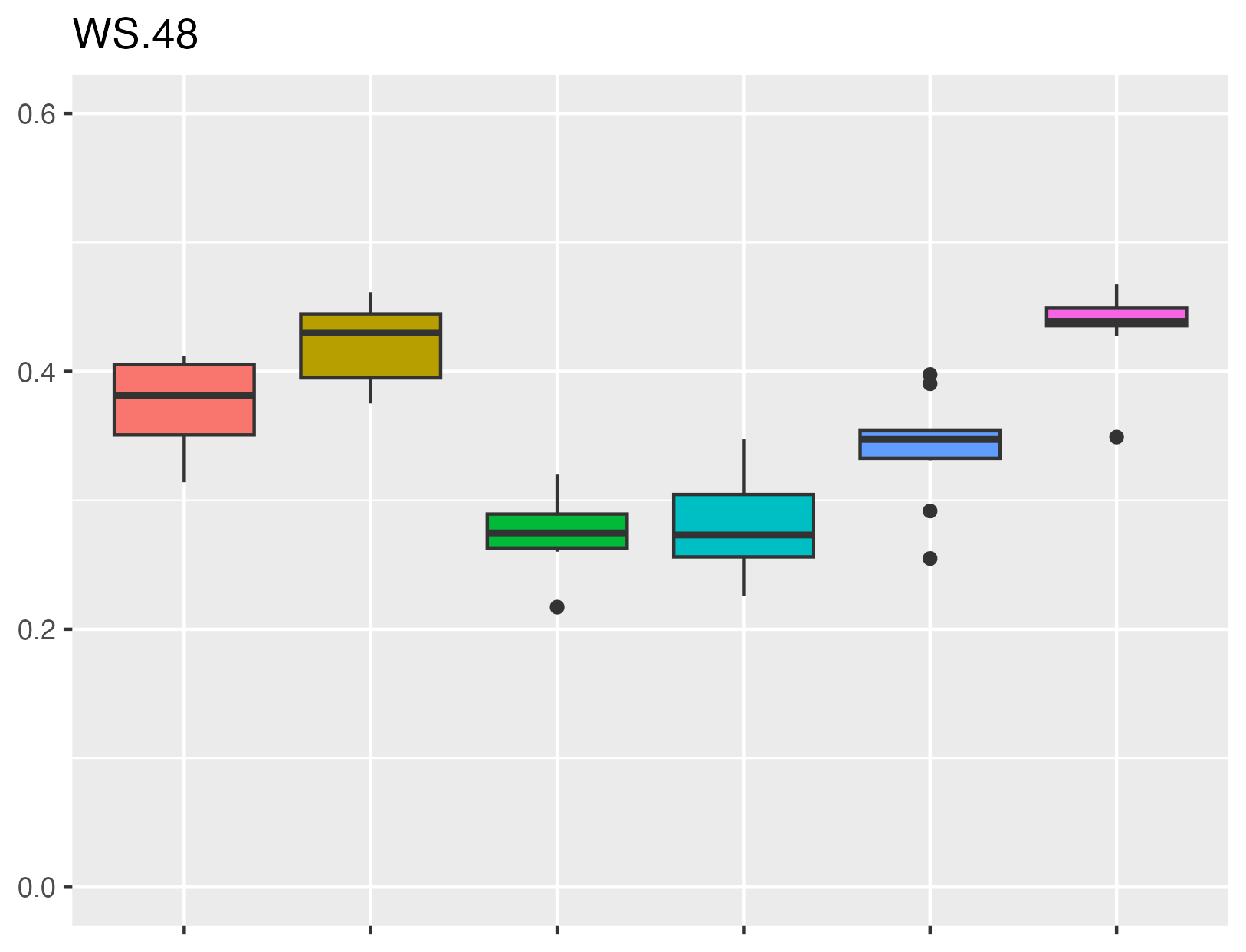}
    \caption{Spearman rank correlation for each method compared to each advanced metric.
    The box-plots show the variability across the years 2012-2022.}
    \label{fig:correlation}
\end{figure}

For each year from 2012-2022, we compute the Spearman rank correlation between each ranking method and the given advanced metric.
The results are presented in Figure~\ref{fig:correlation}.  For methods calculated based on team data (APM, PAPM, HAPM, and LAPM), the fitted values/coefficients across all teams are combined and ranked, so that all correlations are being calculated at the league level. Higher positive correlations imply greater agreement with benchmark rankings, and the box-plots display the variability in the correlations across the years.

In Figure~\ref{fig:correlation}, we see that HAPM is better than APM, i.e. in more agreement with, for BPM, PER, and VORP.
In contrast, LAPM is most highly correlated with WS.48.
This is important because WS.48 considers offensive and defensive contributions (rather than PER, which is primarily a measure of offensive production and scoring efficiency), incorporates playing time, and is highly connected to the corresponding team's overall performance and win record.

We can also see what is lost when moving away from APM\_league, which is modeled at the league level and adjusts for defense.
Moving left to right among the methods, we see that PAPM\_league is generally better than APM\_league for ranking individuals, which shows that adding interaction terms can improve the model's performance.
However, because including additional pair columns in the design matrix requires greater regularization (i.e. shrinkage of all of the coefficients), the improvement is not immense.
We also see that computing APM at the team level is worse than at league level, which makes the performance of HAPM and LAPM even more noteworthy.

From these comparisons, we see that HAPM and LAPM align with advanced metrics as well as APM\_league despite neither explicitly incorporating defensive considerations nor being computed at the league level.
Moreover, HAPM and LAPM offer simultaneous rankings for all generalized lineups and are more robust to choice of penalization parameter.

\subsection{Predictive Accuracy}

\begin{figure}[t!]
    \centering
    \includegraphics[width=.32\textwidth]{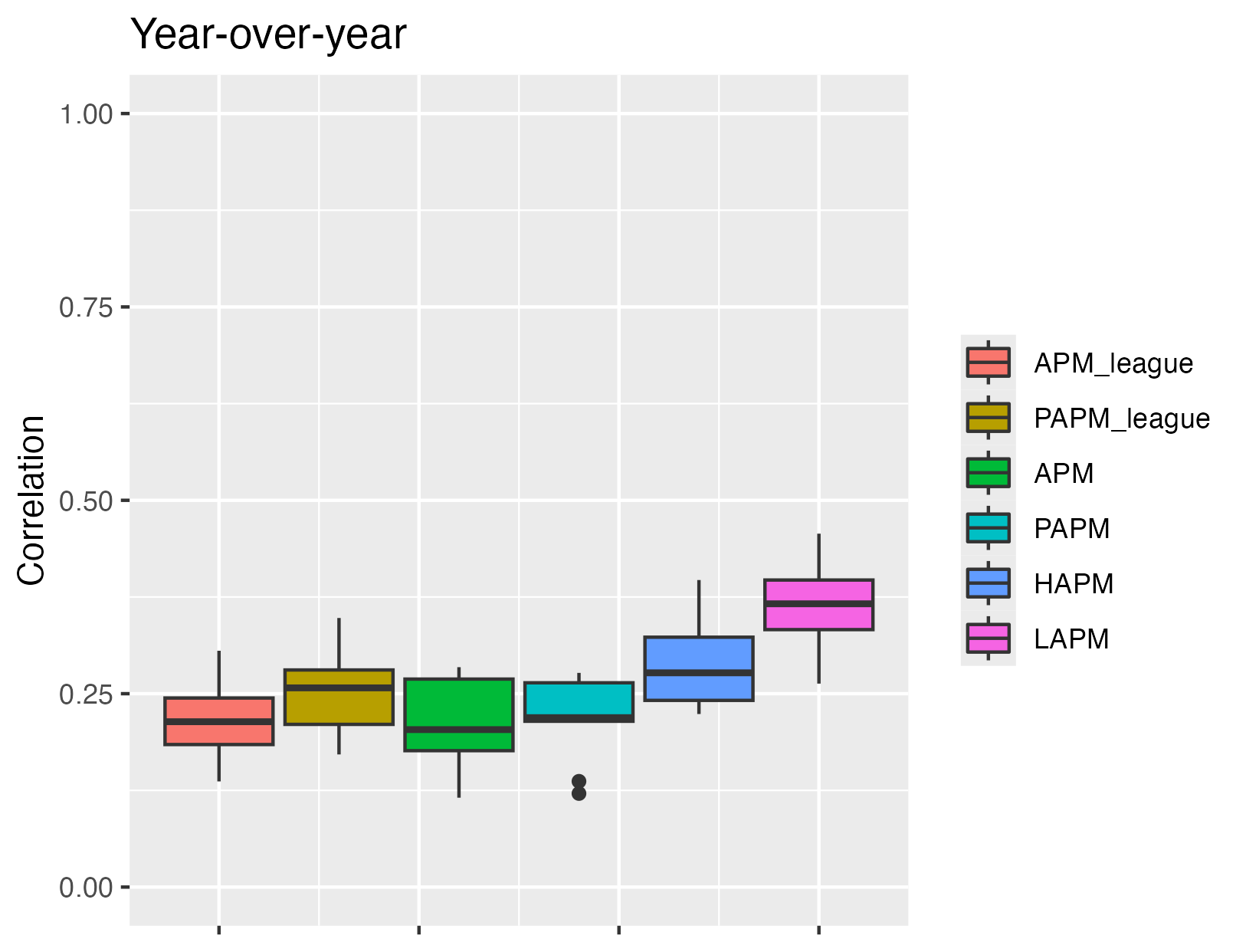}
    \includegraphics[width=.32\textwidth]{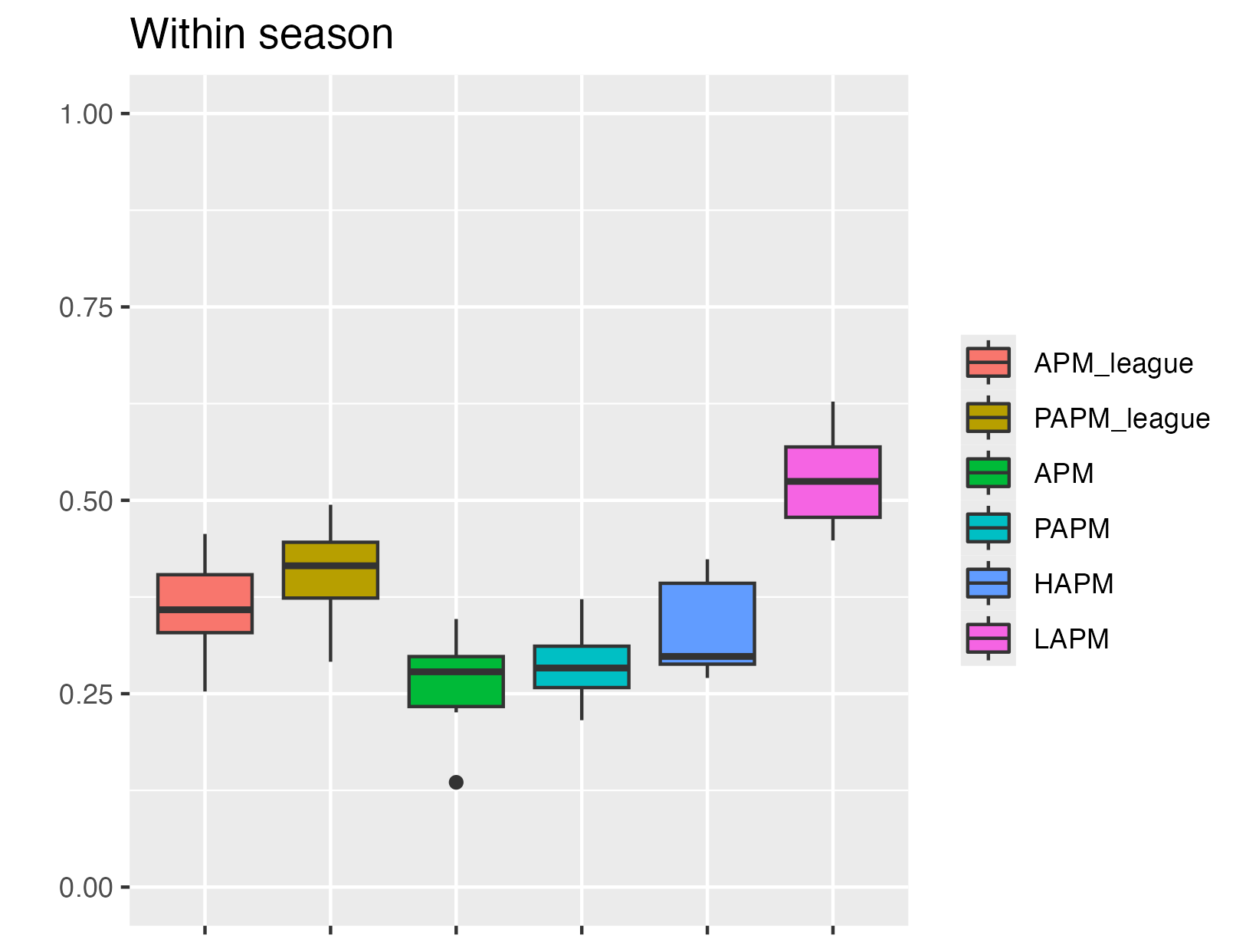}
    \includegraphics[width=.32\textwidth]{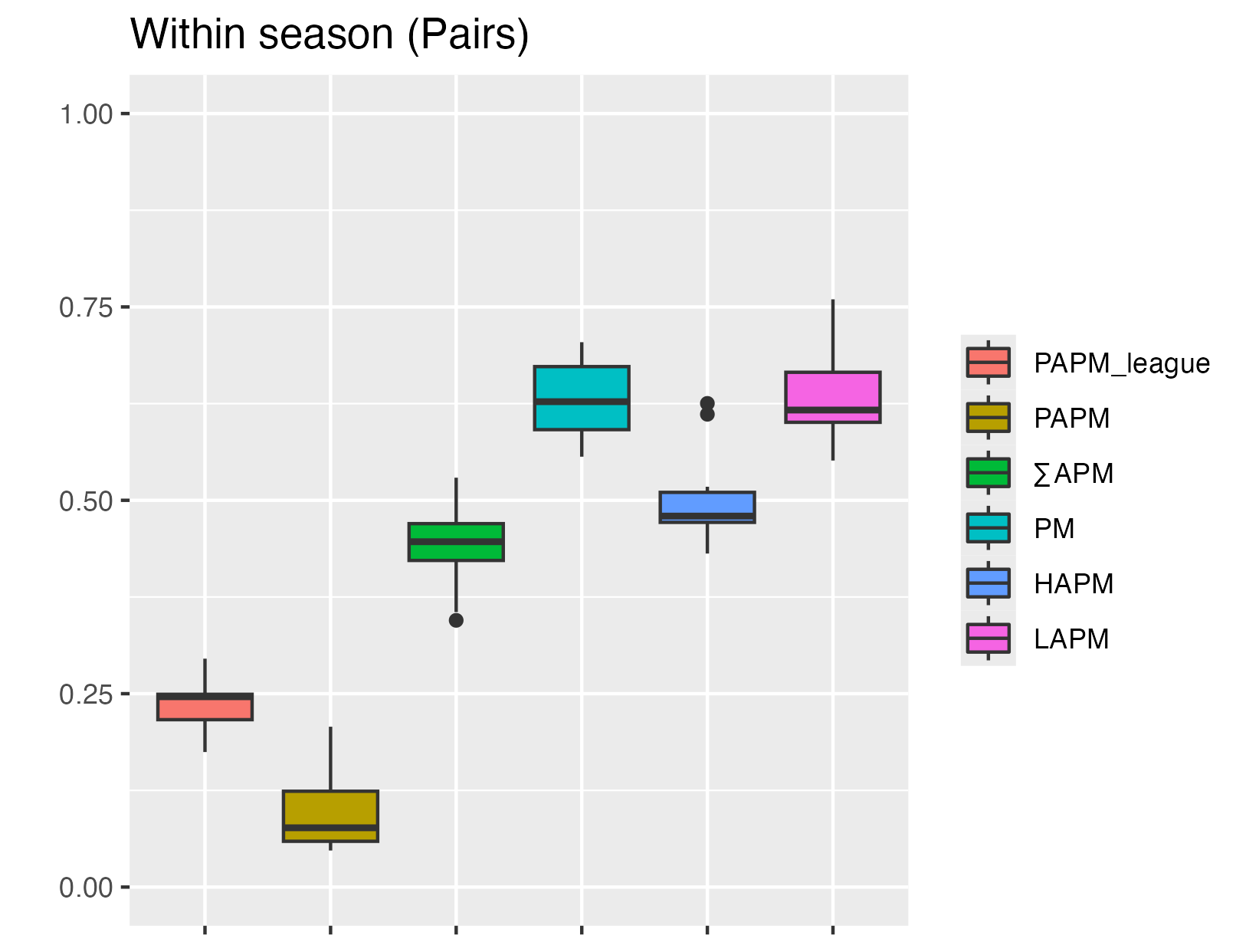}
    \caption{Spearman rank correlation for each method year-over-year (left), within year to observed PM for individuals (middle), and within year to observed PM for pairs (right) across the years 2012-2022.}
    \label{fig:predictive}
\end{figure}

In addition to looking at the agreement between our methods and several advanced analytics, we can see how predictive our metrics are of future player/pair/generalized lineup performance.  
In Figure~\ref{fig:predictive} (left) we display the year-over-year rank correlations of the APM methods for individual players.
For years $t$ and $t+1$, we consider only the players that qualified for inclusion in both years, but we allow those players that switch teams.
We see that HAPM and LAPM have higher correlations than the other APM variants, which suggests HAPM and LAPM are more stable metrics.
However, we note that the rank correlation values are not particularly high, so there is clearly change happening in the rankings over time, which we should expect.

To evaluate within-season prediction capability, we split each season into two halves and assess how well the metrics predict data from the second half of the season when fit on data from the first half of the season.
In Figure~\ref{fig:predictive} (middle), we see that LAPM predictions have the highest rank correlation to the observed, second-half of season, plus-minus values.  However, this evaluation interprets PM as the ground truth, which is somewhat problematic given our starting argument is that PM is not a complete reflection of an individual player's talent.  Therefore, we also include in the appendix (Figure~\ref{fig:predictive_appendix}) the rank correlation between halves of the season for each metric.  This again suggests that HAPM and LAPM are the more stable metrics over time.  That is, incorporating all generalized lineups in the modeling methodology (as HAPM and LAPM do) appears to help player ranking predictions.   

We perform similar evaluations on the ranked pairs.  In addition to PAPM, HAPM, and LAPM, we consider the two naive approaches mentioned in the introduction.
In particular, we let $\Sigma$APM and PM denote the methods that sum the APM of all individuals and the cumulative PM, respectively, for the generalized lineup of interest (in this case, the pairs).  
Again, we see that LAPM is the most predictive method with HAPM also performing well.  
Interestingly, PM is very predictive for pairs, but as we noted, this evaluation method considers PM as the ground truth, so if there are dependencies in, for example, substitution patterns, then the second-half PM values will likely also be affected, which may be artificially inflating the correlation.  As above, we include the correlation between season halves for each metric in Figure~\ref{fig:predictive_appendix}.

\subsection{Case Study}

In this section, we provide a qualitative assessment of our methods by focusing on the Philadelphia 76ers from 2021-22.
This year is notable because the Sixers traded for superstar James Harden midway through the season.
This serves as a robustness check to see if our methods detect Harden's value despite only playing in a fraction of games, and hence having an artificially deflated PM.
Despite his historic scoring prowess, the 2021-22 season continues Harden's transition to a facilitator, so we should expect him to show up in the top pair rankings.
The Sixers roster also includes Joel Embiid, who led the league in scoring and was runner up in MVP voting for the second consecutive year, as well as second-year player Tyrese Maxey who transitioned to a starter and has since become an All Star.

\begin{figure}
    \centering
    \includegraphics[width=.4\textwidth]{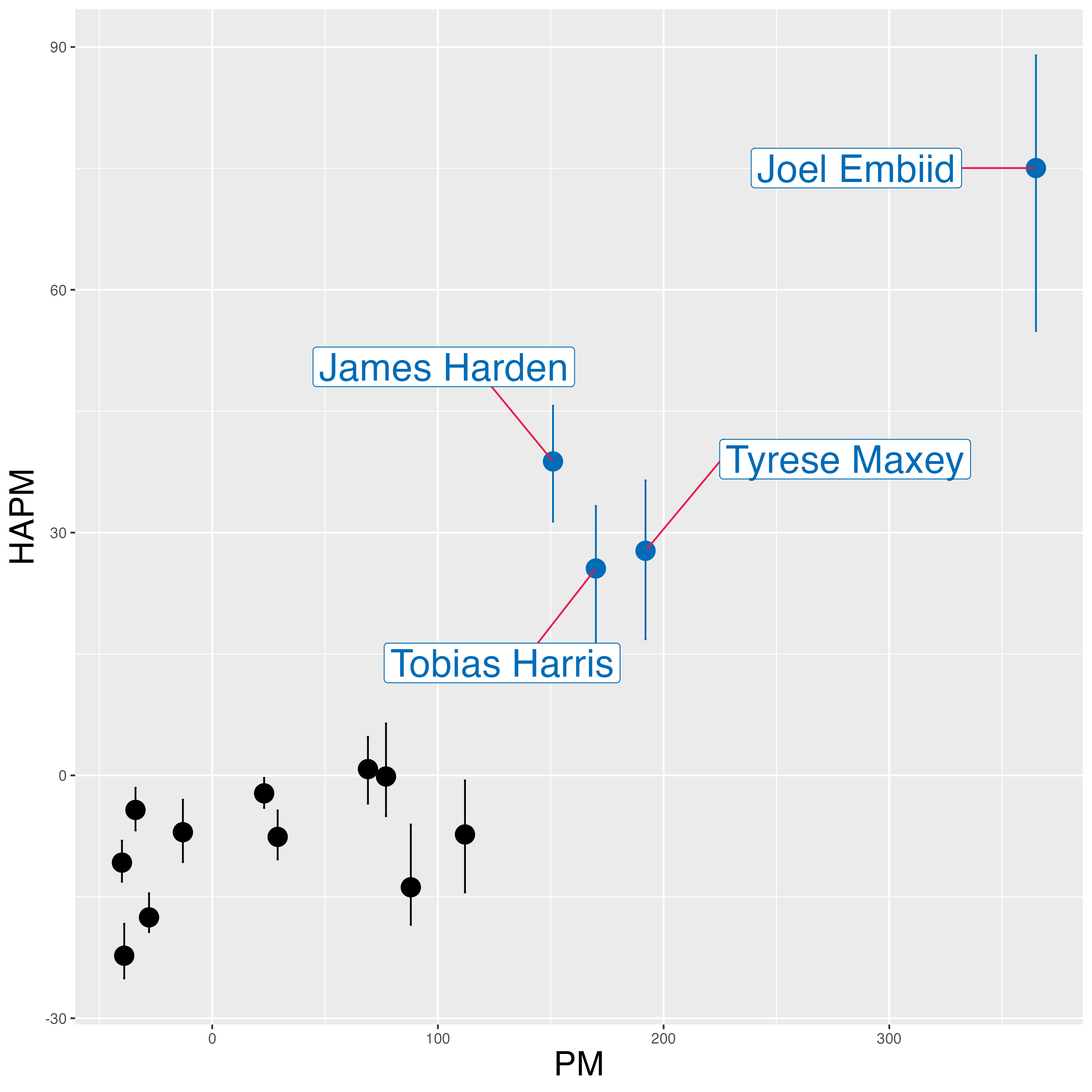}
    \includegraphics[width=.4\textwidth]{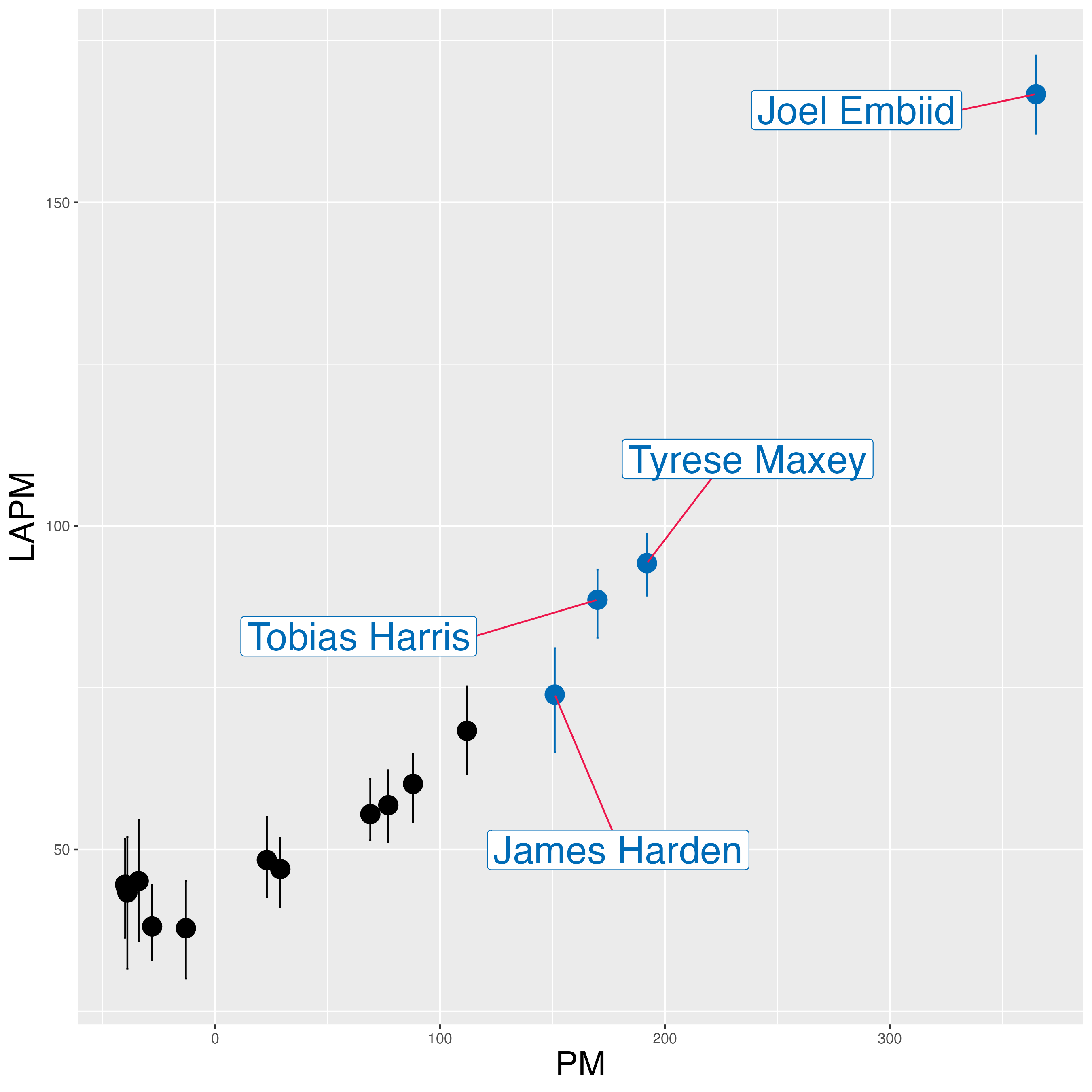}
    \caption{HAPM (left) and LAPM (right) for individual players on the Sixers in 2021-22. Error bars on the y-axes represent inner-quartile range confidence intervals.}
    \label{fig:apm_vs_hapm}
\end{figure}

We show the top eight individual players according to HAPM in Table~\ref{tab:player_data}.
We provide their HAPM and LAPM along with the 50\% confidence intervals for the implied rank.
The full distribution of HAPM/LAPM values is given in Figure~\ref{fig:boxplot_players} in the appendix.
Unsurprisingly, we see the top four players are Embiid, Harden, Maxey, and Tobias Harris, a former first-round draft pick.
Importantly, despite only having a PM of 151, Harden is identified as the second best player by HAPM.
Although there are positive rank correlations between our metrics and PM (0.68 and 0.85 for HAPM and LAPM, respectively), by plotting them against PM in  Figure~\ref{fig:apm_vs_hapm}, we can identify players that may be over or under-valued based on PM alone.

\begin{table}
    \centering
    \begin{tabular}{ccccc}
   
    Player Name & Minutes & PM & HAPM (Rank CI) & LAPM, Rank (Rank CI) \\
    \hline
    Joel Embiid & 2295 & 365 & 75.0 (1 - 1) & 166.7, 1 (1 - 1) \\
    James Harden & 793 & 151 & 38.8 (2 - 3) & 73.9, 4 (4 - 5) \\
    Tyrese Maxey & 2648 & 192  & 27.8 (3 - 4) & 94.2, 2 (2 - 3) \\
    Tobias Harris & 2542 & 170  & 25.6 (3 - 4) & 88.6, 3 (3 - 3) \\
    George Niang & 1742 & 69  & 0.8 (5 - 8) & 55.4, 8 (7 - 9) \\
    Danny Green & 1349 & 77  & -0.1 (5 - 9) & 56.8, 7 (6 - 10) \\
    Isaiah Joe & 609 & 23  & -2.2 (6 - 8) & 48.4,  (8 - 12) \\
    Charles Bassey & 168 & -34  & -4.3 (7 - 10) & 45.1,  (9 - 14) \\
    \end{tabular}
    \caption{Top eight Sixers players in 2021-22 according to HAPM (and 50\% confidence intervals for implied rank).}
    \label{tab:player_data}
\end{table}

\begin{figure}
    \centering
    \includegraphics[width=.4\textwidth]{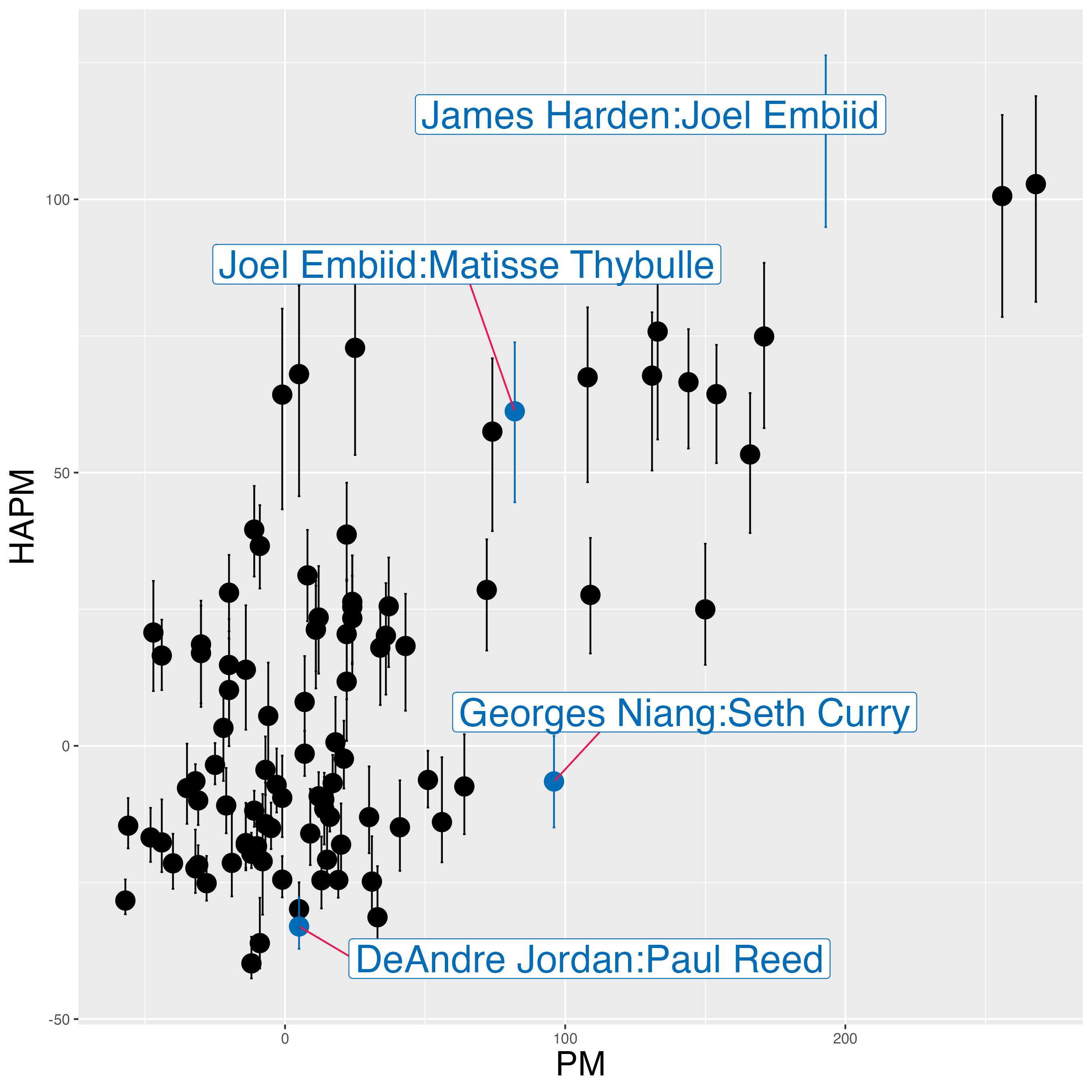}
    \includegraphics[width=.4\textwidth]{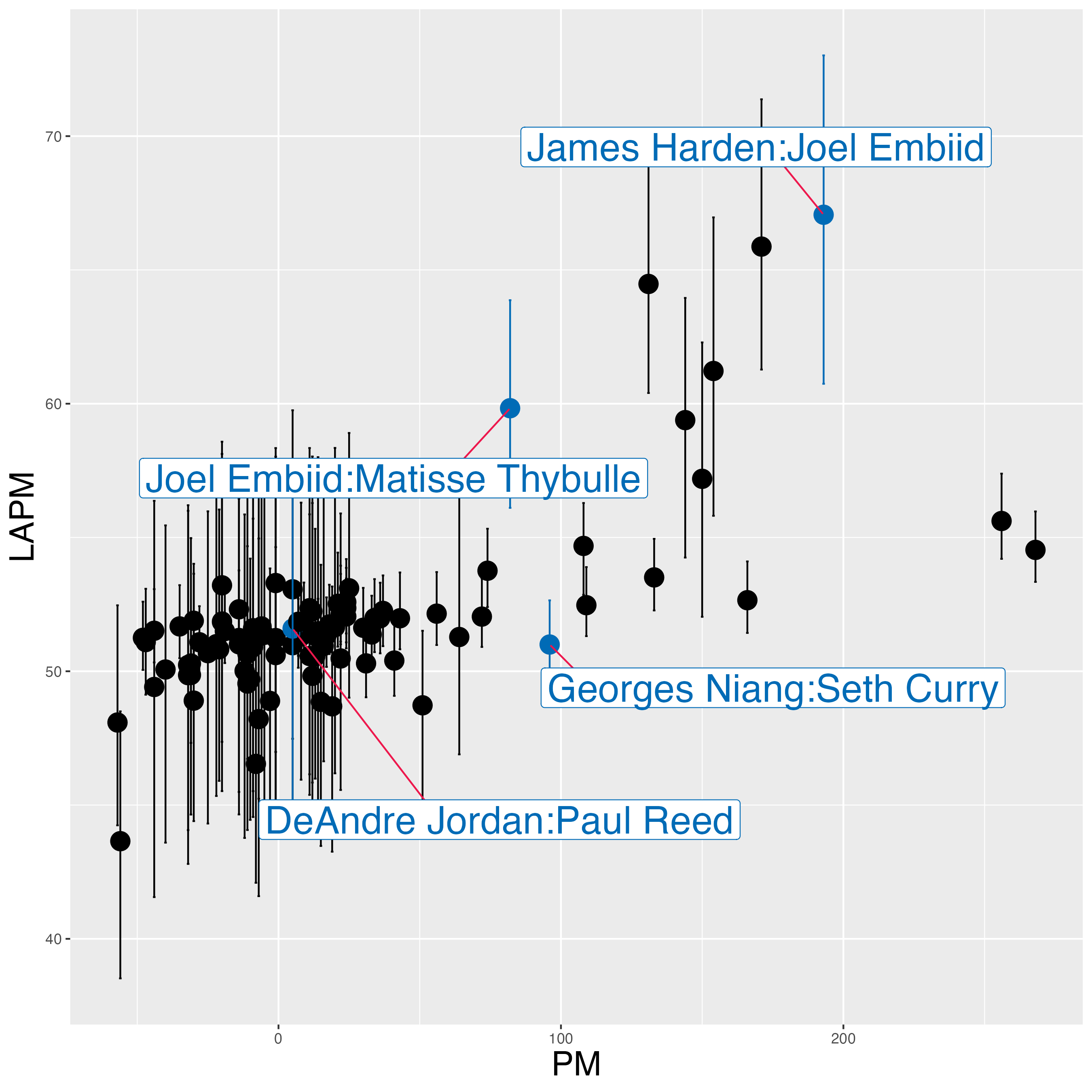}
    \caption{HAPM (left) and LAPM (right) for pairs on the Sixers in 2021-22. Error bars on the y-axes represent inner-quartile range confidence intervals.}
    \label{fig:apm_vs_hapm_pairs}
\end{figure}

We record the top ten pairs according to HAPM in Table~\ref{tab:player_pair_data} and visualize them in Figure~\ref{fig:apm_vs_hapm_pairs}.
The rank correlations with PM are 0.61 and 0.46 for HAPM and LAPM, respectively.
Here we see that not only is Embiid a great individual player, but he contributes to the top nine pairings according to HAPM (and seven of the top ten best pairings according to LAPM).
This is consistent with Embiid's reputation as a player with the ability to pass out of a double team.
Furthermore, despite only playing together for 600 total minutes, both HAPM and LAPM recognize the successful pairing of Embiid and Harden.
As we hoped, our metrics identify Harden as a good teammate, belonging to three of the top six pairs for both HAPM and LAPM.

\begin{table}
    \centering
    \begin{tabular}{cccc}
  
    Player Pair & PM & HAPM (Rank CI)  & LAPM, Rank, (Rank CI) \\
    \hline
    
    James Harden:Joel Embiid & 193 & 113.8 (1 - 2) & 67.1, 1 (2 - 9) \\
    Joel Embiid:Tyrese Maxey & 268 & 102.8 (1 - 3) & 54.5, 10 (20 - 26) \\
    Joel Embiid:Tobias Harris & 256 & 100.6 (2 - 4) & 55.6, 8 (17 - 23) \\
    Georges Niang:Joel Embiid & 133 & 75.8 (5 - 9) & 53.5, 12 (26 - 33) \\
    Danny Green:Joel Embiid & 171 & 74.9 (5 - 9) & 65.9, 2 (2 - 9) \\
    Isaiah Joe:Joel Embiid & 25 & 72.8 (6 - 9) & 53.1, 15 (13 - 78) \\ 
    Andre Drummond:Joel Embiid & 5 & 68.0 (7 - 14) & 53.1, 16 (13 - 79) \\
    Joel Embiid:Seth Curry & 131 & 67.7 (8 - 12) & 64.5, 3 (3 - 10) \\
    Furkan Korkmaz:Joel Embiid & 108 & 67.4 (8 - 13) & 54.7 (19 - 28) \\
    James Harden:Tyrese Maxey & 144 & 66.5 (4 - 13) & 59.4, 6 (6 - 24) 
    \end{tabular}
    \caption{Top ten Sixers pairs in 2021-22 according to HAPM (and 50\% confidence intervals for implied rank).}
    \label{tab:player_pair_data}
\end{table}

Interestingly, we see high rankings for the pairings of Embiid and Andre Drummond (9th best pair according to HAPM and 16th best pair according to LAPM, out of 96 qualifying pairs) and Embiid and Matisse Thybulle (13th/5th according to HAPM/LAPM) despite having PMs of only 5 and 82, respectively.
This can be explained by the excellent defensive pairing of Embiid with Drummond and Thybulle, which both of our methods are able to detect.
Not surprisingly, one of the worst ranking pair according to HAPM (shown in Figure~\ref{fig:apm_vs_hapm_pairs}) is Deandre Jordan and Paul Reed, who are both big men with overlapping skillsets.
It is also interesting that the two best pairs according to PM (Embiid and Maxey \& Embiid and Harris) have relatively low pair rankings (10th and 8th, respectively) according to LAPM.
This may suggest that the high PM of these pairings can be attributed to individual player success or the success of higher-order generalized lineups (e.g. Embiid and Harris play together in the 3rd, 9th, 13th and 14th best ranked trios according to LAPM).

\begin{figure}[]
    \centering
    \includegraphics[width = .49\textwidth]{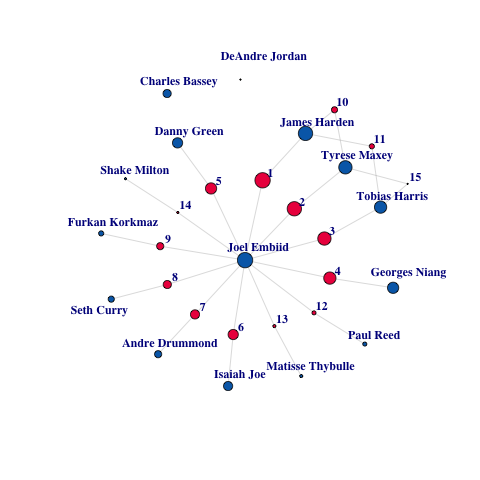}
    \includegraphics[width = .49\textwidth]{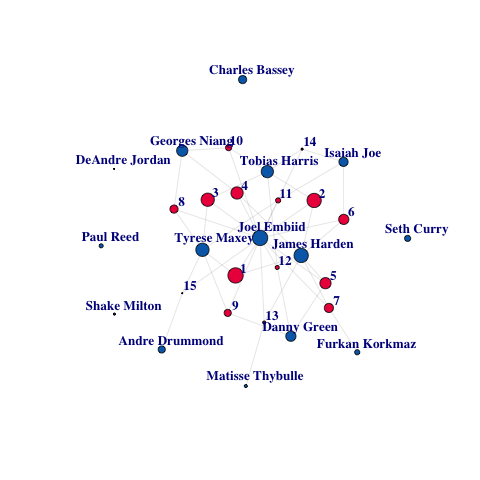}
    \caption{Line graph visualization of player/pair (left) and player/trio (right)  HAPM rankings for PHI 2021-22.  The size of the nodes corresponds to ranking within the individuals/pairs/trios.
    The blue nodes represent players, and the red nodes, labeled with rank, represent the pair/trio.}
    \label{fig:line_graph}
\end{figure}

Finally, we show that the connection between HAPM, LAPM, and graphs can be leveraged to create informative visualizations.
Figure~\ref{fig:line_graph} displays the line graphs based on the top 15 pairs (left) and trios (right) for HAPM.
The blue nodes represent the individual players, while the red nodes correspond to pairs or trios.  The nodes are sized proportional to their HAPM, and an edge exists between the player and pair/trio node if that player belongs to the given generalized lineup.
For example, the best pair (labeled 1) is connected to the Embiid and Harden nodes, whereas the best trio is connected to Embiid, Harden, and Maxey.
The layout of the nodes is determined by the Fruchterman-Reingold algorithm, which aims to automatically  produce visually appealing network layouts by distributing nodes evenly and minimizing edge crossings.

An immediate takeaway is the central role of Embiid, as he is part of 12 of the best 15 pairs and all of the 15 best trios according to HAPM.
In both visualizations, he is at the center of the network.  Exploratory analyses for other teams and other years reveals that this is very common; the superstars are the most central nodes in the visualizations (also shown for the Boston Celtics in Figure~\ref{fig:bos_line_graph} in the appendix).
Further investigation reveals that pair nodes that are larger than one or both of the player nodes (Embiid and Danny Green; Embiid and Thybulle; Embiid and Seth Curry; Embiid and Shake Milton) may represent possible synergies, whereas pair nodes that are smaller than the player nodes (Embidd and Isaiah Joe; Harris and Maxey) might indicate inefficient pairings.

\section{Discussion}\label{sec:discussion}

In this paper, we propose a new framework for ranking individual players using plus-minus data, while simultaneously ranking all \textit{genearlized lineups}.
We show that our methods, HAPM and LAPM, are similarly or more correlated with advanced metrics for ranking individual players than ordinary APM.
We also provide a case study on the Philadelphia 76ers in 2021-22 that seemingly identifies the best individuals, pairs, and trios.  In terms of utility, coaches can identify under- and over performing player combinations mid-season, adjust their lineups accordingly, and evaluate the impact on team performance. Moreover, network visualizations illustrate these player interactions, directly highlighting which combinations drive success or need improvement.

There a few important things to note about our methods.
First, as previously mentioned, PM, APM, and its variants including our proposed HAPM and LAPM are highly correlated with a team's winning percentage.
This is not inherently bad since winning is important, but this is also not surprising because these methods only use plus-minus data, whereas advanced metrics use many covariates.
HAPM and LAPM provide an improvement over APM with respect to individual rankings, as measured by correlation to advanced metrics, \textit{while only using plus-minus data}.

There is room for refinement of our methods, as some decisions in our data processing were done out of convenience.
For example, we did not track PM exactly for substitutions that occur during free throws.  We also chose Jaccard similarity to define the weights for the line graph of the hypergraph, but other similarity measures could be investigated.  In terms of model fitting, we used ridge regression for simplicity and comparison to commonly published basketball metrics, but did not think deeply about choosing $\lambda$.  Similarly, with LAPM, a more thorough investigation of the connection between the basis rank expansion (captured by $\tau$) and $\lambda$ may lead to improved predictive capabilities.

There are several directions for future research.  Here, we presented two new metrics, LAPM and HAPM, for ranking generalized lineups.  While we compared their results and stated the theoretical differences in the two approaches, we did not attempt to understand the cause of the metrics (dis)agreement.
This may be a case-by-case exercise that could shed light on the inner workings of any one team.  Furthermore, while it is straightforward to extend HAPM and LAPM to other APM variants that account for other non box-score statistics \citep{kubatko2007starting}, it is less obvious how to incorporate defense for HAPM and LAPM.  Our work demonstrates that HAPM and LAPM can achieve comparable results to metrics calculated at the league level even without incorporating defense, but this is a natural future direction. 





\bibliography{refs}
\bibliographystyle{agsm}

\pagebreak
\appendix

\setcounter{equation}{0}
\setcounter{figure}{0}
\setcounter{table}{0}
\setcounter{page}{1}
\setcounter{section}{0}
\makeatletter
\renewcommand{\theequation}{S\arabic{equation}}
\renewcommand{\thefigure}{A\arabic{figure}}
\renewcommand{\bibnumfmt}[1]{[A#1]}
\renewcommand{\citenumfont}[1]{A#1}

\section{Appendix}\label{sec:appendix2}

In the LAPM model specification, we have
\begin{align*}
    Y_v~|~\beta &\stackrel{\text{ind}}{\sim} N(\beta_v, ~\sigma^2 ) \\
    \beta &\sim N(0, ~\lambda^{-1}L_w^{-1}).
\end{align*}
To evaluate the uncertainty in our modeling process, one option is to fix a value for $\sigma^2$, i.e. view it as a known value.
Alternatively, we could extend the model definition by allowing $\sigma^2$ to be a random variable with an inverse scaled chi-squared prior:
\begin{align*}
    Y_v~|~\beta &\stackrel{\text{ind}}{\sim} N(\beta_v, ~\sigma^2 ) \\
    \beta &\sim N(0, ~\lambda^{-1}L_w^{-1}) \\
    \frac{1}{\sigma^2} &\sim \Gamma (\nu/2, \frac{\nu \tau^2}{2}).
\end{align*}

For the analysis presented in the paper, we chose to use a non-informative prior for $\sigma^2$, setting $\nu = \tau^2 = 0.$ We utilized the Metropolis adjusted Langevin algorithm~\citep{girolami} to sample from the posterior distributions.
The posterior distributions displayed throughout the manuscript are based on 1000 samples after disregarding 10\% for a burn-in period and thinning every 5 samples.  

We note that while the center and spread of the posterior distributions (and therefore, the fitted values) are sensitive to prior specification, the resulting \textit{rankings} of the generalized lineups are not.
Also, while we set the hyper-prior parameters to be the same for all teams, it may be worthwhile to investigate the parameters at the team level if that is where one's interests primarily lie.   

\section{Appendix}\label{sec:appendix3}

\begin{figure}[H]
    \centering
    \includegraphics[width = .49\textwidth]{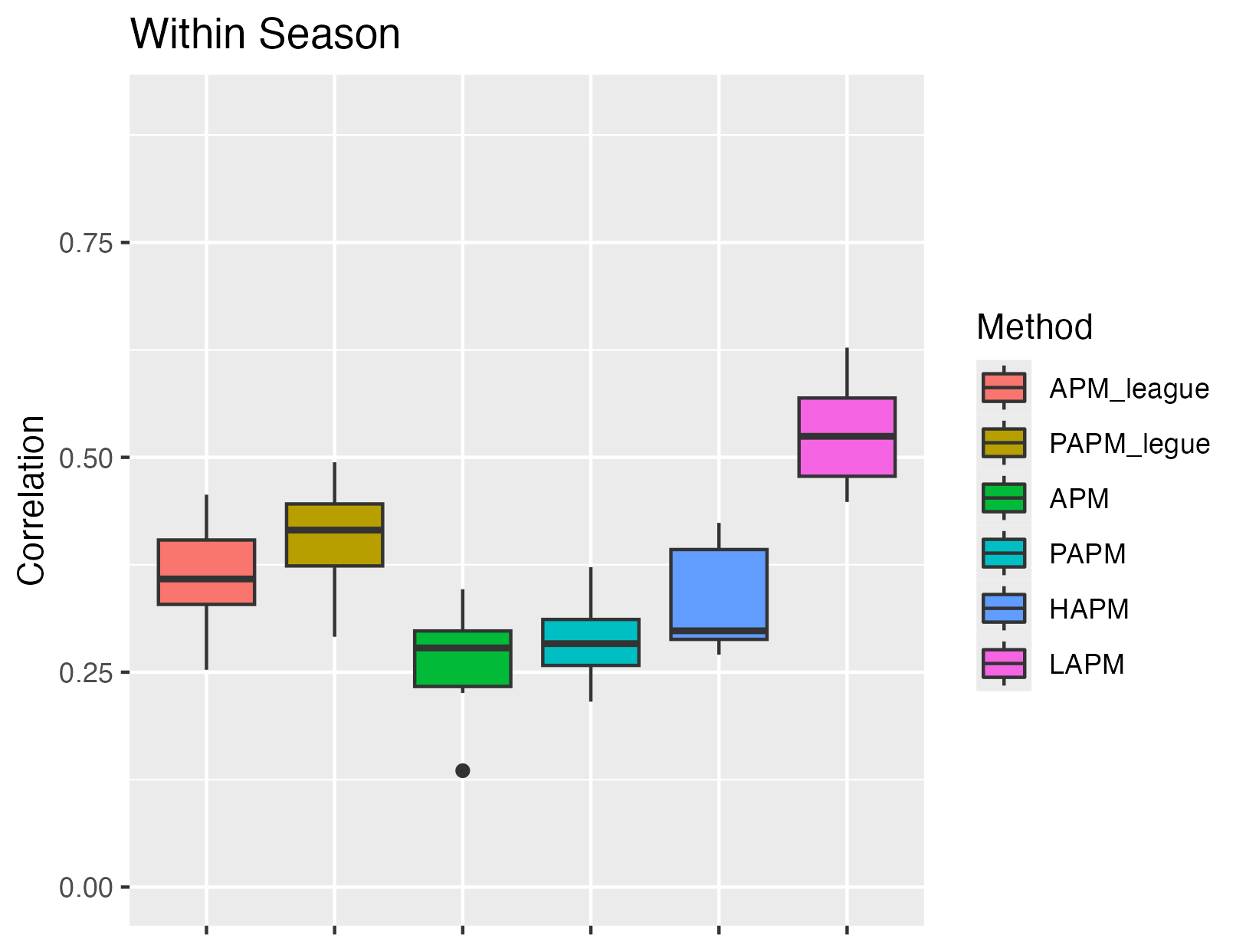}
    \includegraphics[width = .49\textwidth]{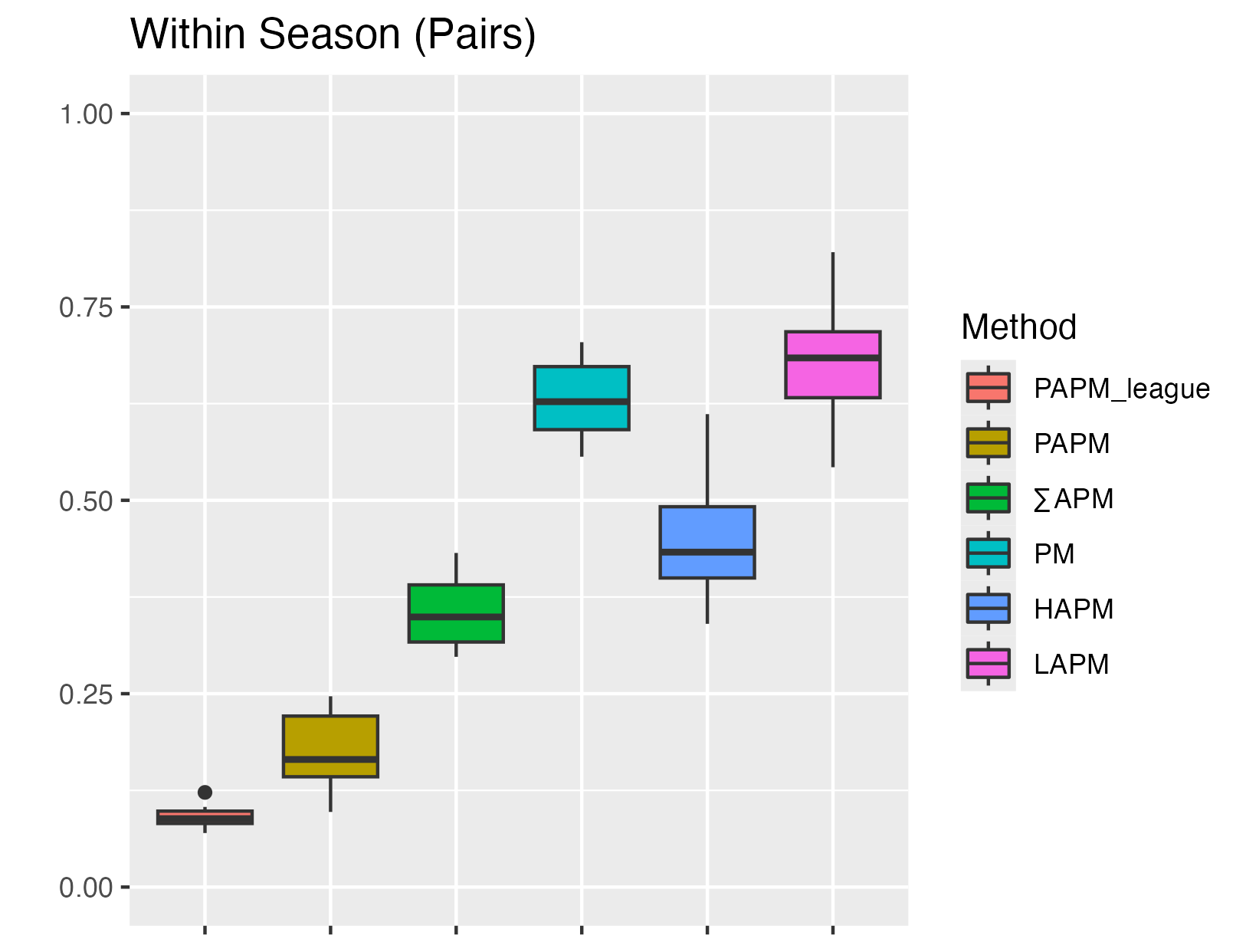}
    \caption{Spearman rank correlation for each method within year to itself for individuals (left) and pairs (right) across the years 2012-2022.}
    \label{fig:predictive_appendix}
\end{figure}

\section{Appendix}\label{sec:appendix1}

\begin{figure}[H]
    \centering
    \includegraphics[width = .49\textwidth]{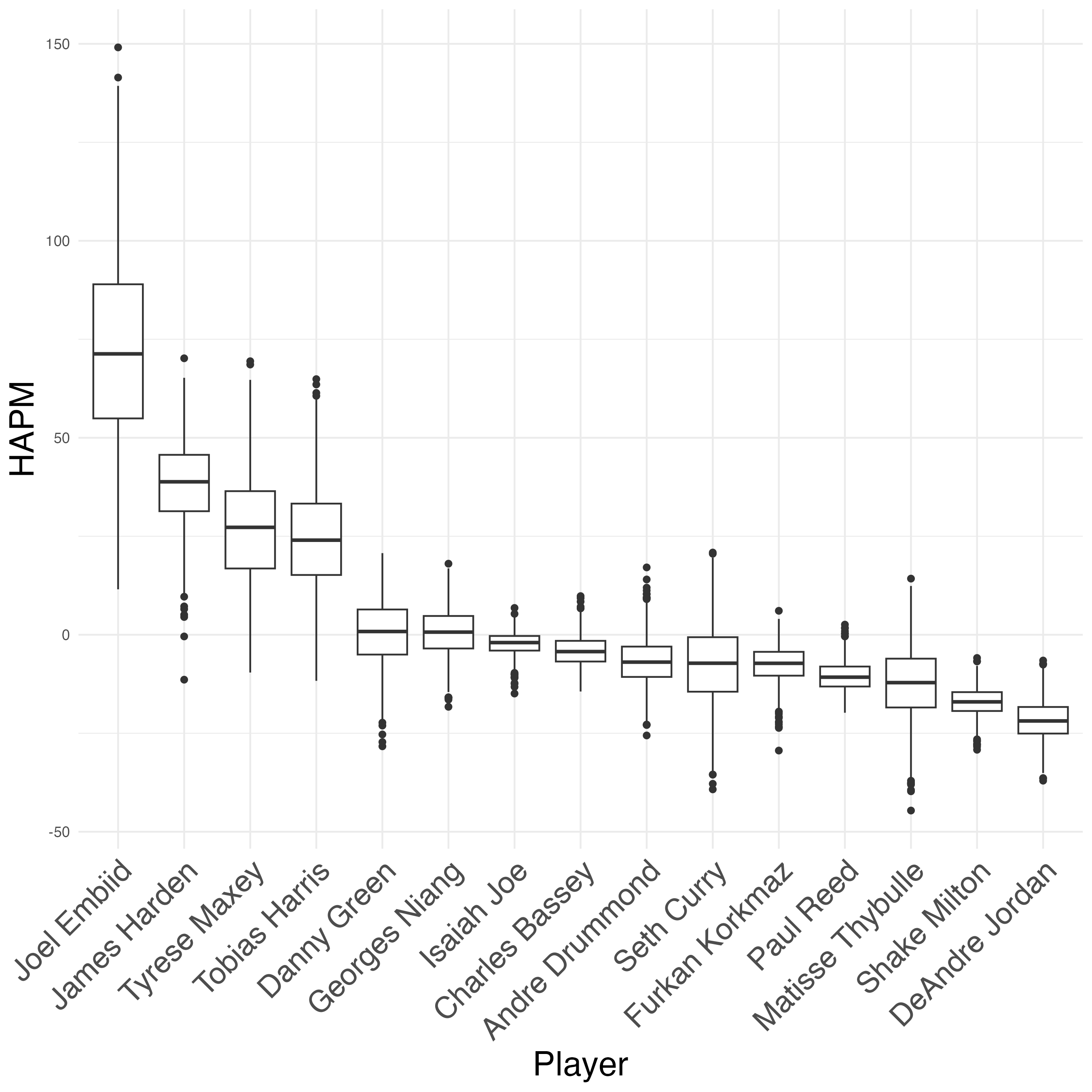}
    \includegraphics[width = .49\textwidth]{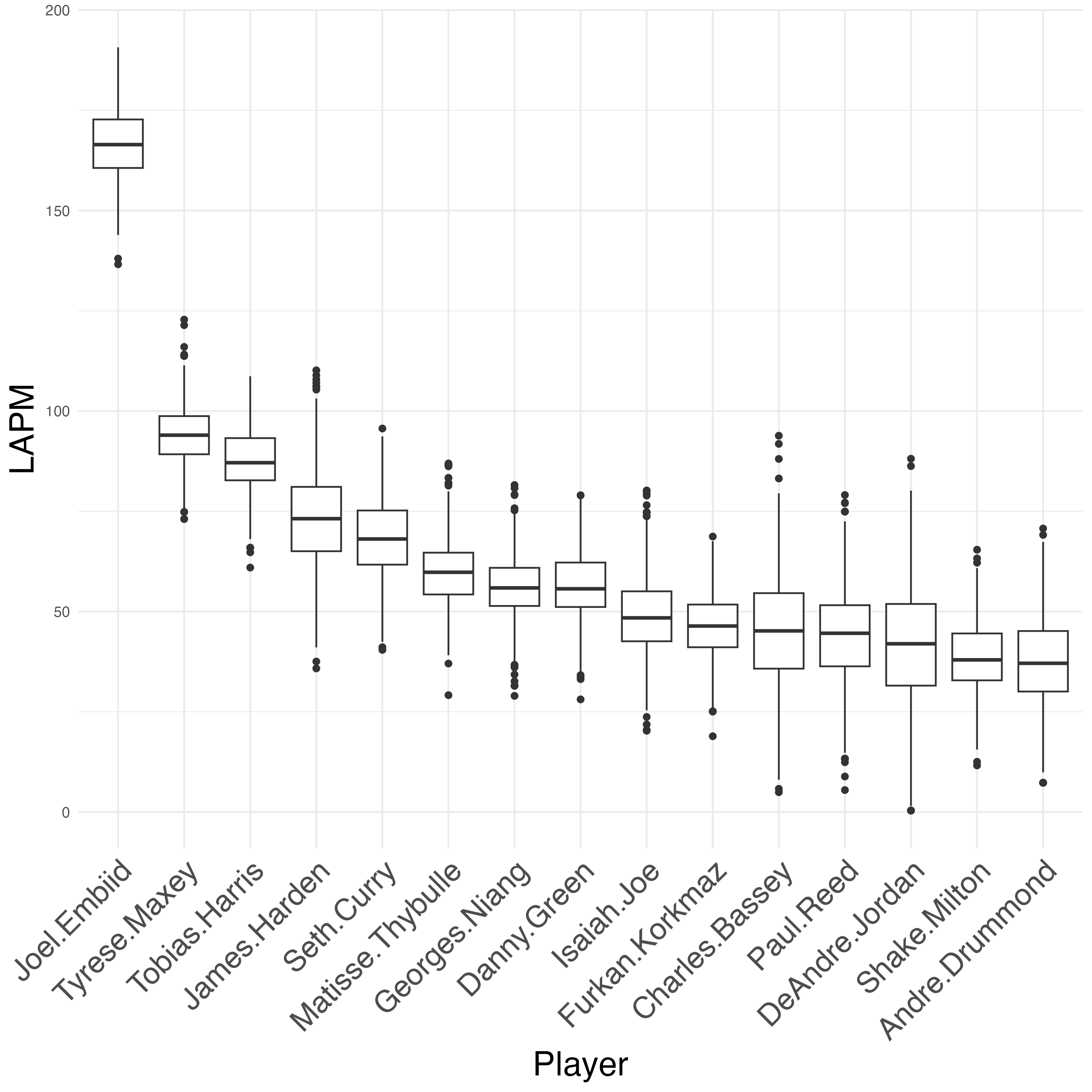}
    \caption{HAPM (left) and LAPM (right) for individuals  on the Sixers in 2021-22. The boxplots represent the distribution of HAPM and LAPM values obtained via bootstrap and MCMC, respectively.}
    \label{fig:boxplot_players}
\end{figure}

\begin{figure}[H]
    \centering
    \includegraphics[width = .49\textwidth]{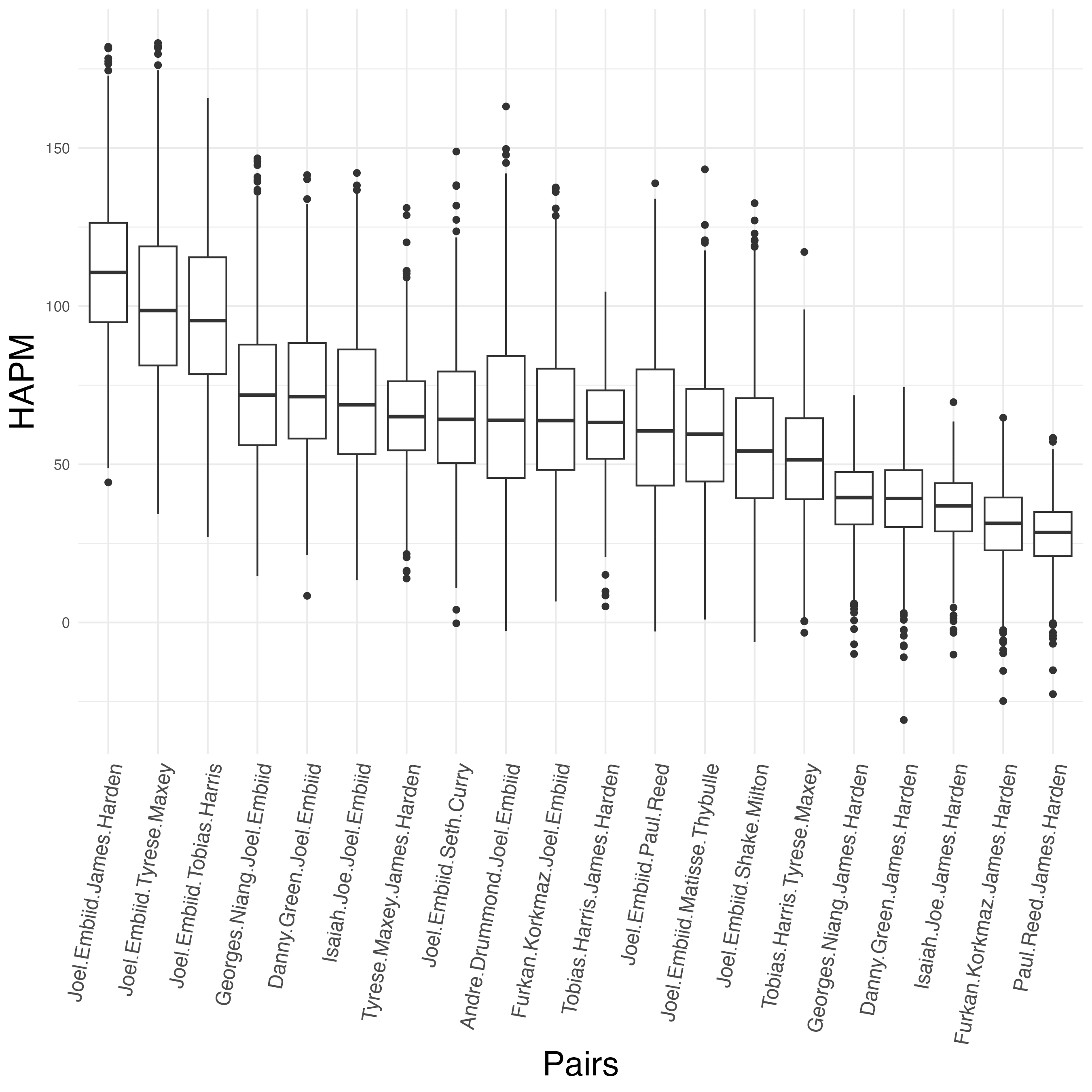}
    \includegraphics[width = .49\textwidth]{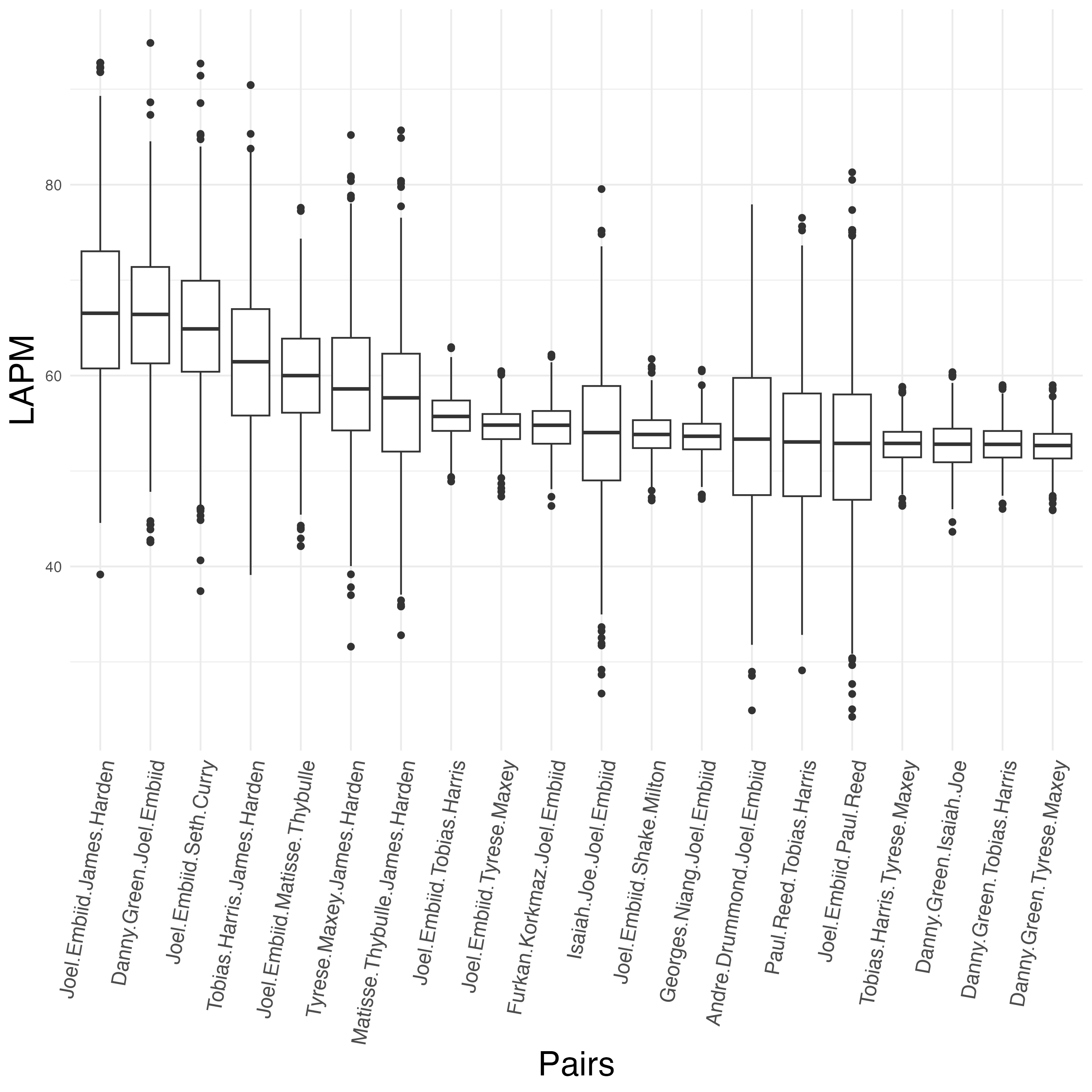}
    \caption{HAPM (left) and LAPM (right) for the top pairs  on the Sixers in 2021-22. The boxplots represent the distribution of HAPM and LAPM values obtained via bootstrap and MCMC, respectively.}
    \label{fig:boxplot_pairs}
\end{figure}

\begin{figure}[]
    \centering
    \includegraphics[width = .49\textwidth]{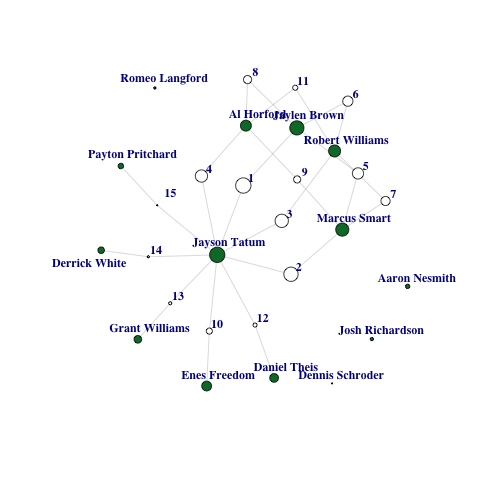}
    \includegraphics[width = .49\textwidth]{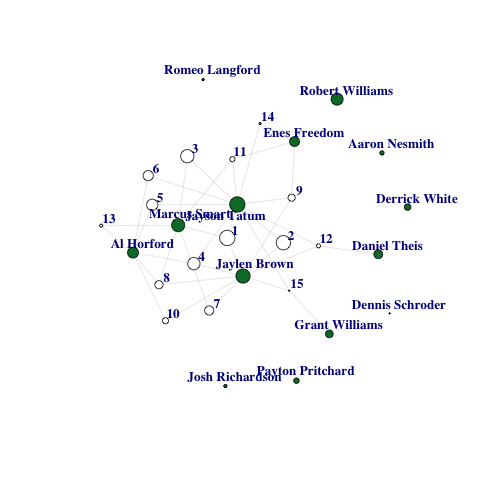}
    \caption{Line graph visualization of player/pair (left) and player/trio (right)  HAPM rankings for BOS 2021-22.  The size of the nodes corresponds to ranking within the individuals/pairs/trios.
    The green nodes represent players, and the white nodes, labeled with rank, represent the pair/trio.}
    \label{fig:bos_line_graph}
\end{figure}

\end{document}